\documentclass{interact}
\usepackage[longnamesfirst,sort]{natbib}
\bibpunct[, ]{(}{)}{;}{a}{,}{,}
\renewcommand\bibfont{\fontsize{10}{12}\selectfont}
\usepackage[normalem]{ulem}
\useunder{\uline}{\ul}{}
\usepackage{caption}
\usepackage{longtable}
\usepackage{graphicx}
\usepackage{booktabs}
\usepackage{colortbl}
\usepackage{subfloat}
\usepackage{amsmath}
\usepackage{amssymb}
\usepackage{epigraph}

\makeatletter
\newlength\epitextskip
\usepackage{hyperref}
\usepackage{xcolor}
\hypersetup{
            colorlinks = true,
            linkcolor=blue,
             urlcolor=blue,
             citecolor=blue,
            allcolors = blue,
            linkbordercolor = {white}}
\usepackage{pgf-pie}
\usetikzlibrary{angles,calc,positioning}
\theoremstyle{plain}

\theoremstyle{definition}

\theoremstyle{remark}

\usepackage{fancyhdr}
\usepackage{orcidlink}
\usepackage{graphicx}
\usepackage{subcaption}
\usepackage{makecell}
\usepackage{float}

\usepackage{comment}

\begin{document}

\title{Collusion Detection with Graph Neural Networks}          
\author{
      \name{Lucas Gomes\textsuperscript{a}, 
      Jannis Kueck\textsuperscript{b},
      Mara Mattes\textsuperscript{b}, 
      Martin Spindler\textsuperscript{a}, 
      Alexey Zaytsev\textsuperscript{c}} 
    \affil{\textsuperscript{a}University of Hamburg, Faculty of Business Administration}
    \affil{\textsuperscript{b}Heinrich Heine University Düsseldorf, Düsseldorf Institute for Competition Economics}
    \affil{\textsuperscript{c}Skolkovo Institute of Science and Technology}
    }

\maketitle
             
    \begin{abstract}

 Collusion is a complex phenomenon in which companies secretly collaborate to engage in fraudulent practices. This paper presents an innovative methodology for detecting and predicting collusion patterns in different national markets using neural networks (NNs) and graph neural networks (GNNs). GNNs are particularly well suited to this task because they can exploit the inherent network structures present in collusion and many other economic problems. Our approach consists of two phases: In Phase I, we develop and train models on individual market datasets from Japan, the United States, two regions in Switzerland, Italy, and Brazil, focusing on predicting collusion in single markets. In Phase II, we extend the models’ applicability through zero-shot learning, employing a transfer learning approach that can detect collusion in markets in which training data is unavailable. This phase also incorporates out-of-distribution (OOD) generalization to evaluate the models’ performance on unseen datasets from other countries and regions. In our empirical study, we show that GNNs outperform NNs in detecting complex collusive patterns. This research contributes to the ongoing discourse on preventing collusion and optimizing detection methodologies, providing valuable guidance on the use of NNs and GNNs in economic applications to enhance market fairness and economic welfare.

\end{abstract}


    \section{Introduction}
    
        Collusion detection is a critical process for identifying instances in which entities, such as companies or individuals, secretly collaborate to engage in fraudulent practices, often characterized by agreements on prices, market shares, or tactics to avoid competition \citep{rodriguez2022collusion}. Because such anti-competitive behavior can lead to market manipulation, reduce consumer welfare, and suppress innovation, effective collusion detection is key to maintaining market integrity \citep{10.1093/joclec/nhaa031}. According to the OECD\footnote{OECD: Organisation for Economic Co-operation and Development. \hyperlink{https://www.oecd.org/competition/cartels/fightingbidrigginginpublicprocurement.htm}{oecd.org}}, the average cost of procurements is estimated to be 20\% higher due to collusive practices. 

In the public policy and administration sector, collusion is a major concern due to the strong financial incentives for companies to engage in such practices and the substantial damage they cause. Various countries have established regulatory bodies to address these challenges, such as the CADE\footnote{CADE: Administrative Council for Economic Defense. \hyperlink{http://www.cade.gov.br/}{cade.gov.br}.} in Brazil, the BKartA\footnote{BKartA: Bundeskartellamt.  \hyperlink{https://www.bundeskartellamt.de/}{bundeskartellamt.de}.} in Germany, the FTC\footnote{FTC:  Federal Trade Commission. \hyperlink{https://www.ftc.gov/}{ftc.gov}.} in the United States (US), the COMCO\footnote{COMCO: Competition Commission of Switzerland.  \hyperlink{https://www.weko.admin.ch}{weko.admin.ch}.}  in Switzerland, the AGCM\footnote{AGCM: Italian Competition Authority.  \hyperlink{https://en.agcm.it/en/}{en.agcm.it}.} in Italy, the FAS\footnote{FAS: Federal Antimonopoly Service.  \hyperlink{http://fas.gov.ru/}{fas.gov.ru}.} in Russia and the JFTC\footnote{JFTC: Japan Fair Trade Commission.  \hyperlink{https://www.jftc.go.jp/en/index.html}{jftc.go.jp}.} in Japan. To detect and prevent cartel activities, these entities employ a variety of strategies, ranging from sophisticated econometric models to simpler methods such as analyzing average bid values in auctions. For example, in the average bid auction (ABA) system, an increase in average prices can be associated with the participation of specific groups based on similarity, allowing thresholds to be defined that can be used to classify collusive companies \citep{conley2016detecting}.
Each country's regulatory framework reflects its unique legal and economic context, but all share the common goal of promoting fair competition and protecting consumers from the detrimental effects of collusive practices \citep{arena201760}.

Previous studies have compared the effectiveness of various machine learning techniques in collusion detection with mixed results. In particular, these models, due to an overall imbalance in classification, tend to have a relatively high rate of false negatives \citep{rodriguez2022collusion}.  
One promising approach to improving collusion detection involves the use of modern machine learning methods, especially deep learning architectures. Unlike traditional methods that rely on manually selecting and engineering features (variables) predictive of collusion, deep learning models use a data-driven approach to automatically select and generate relevant features through complex, non-linear transformations in hidden layers. 
Theory-driven models often use a priori knowledge to create screening variables, transforming the original input to include metrics such as distance, which are then used as input variables. Recent studies have explored different machine learning models using the screening variables strategy and have reported improved performance \citep{wallimann2022machine, rodriguez2022collusion, imhof2021detecting}.

Feedforward neural networks (NNs) have also demonstrated effectiveness in modeling latent variables to identify relationship patterns in data, achieving good results when sufficient data is available \citep{wu2020comprehensive,huber2023flagging}. However, feedforward NNs usually assume independence among observations, which represents a substantial limitation in the context of collusion detection because the previous and next bids from the same company, as well as bids from other companies, are strongly associated. 
Modeling these associations with feedforward NNs can lead to overfitting and feature dependency, resulting in lower-quality models. 

\cite{huber2023flagging} have successfully applied convolutional neural networks (CNNs) to graphical plots of bid interactions, yielding promising results in classifying collusive companies when applying their model to Japanese and Swiss bid data, as well as when evaluating their Japanese model on the Swiss dataset,
and their Swiss model on the Japanese dataset. 

Graph neural networks (GNNs) are graph-structured models that allow information to flow among neighboring nodes through a message-passing mechanism. This approach allows for observation dependence, facilitating the sharing of information, such as from previous bids to subsequent bids from the same company, as well as between bids from different companies involved in the same tender (network structure). GNNs can also handle different types of connections by assigning distinct weights to each, thus preserving the meaning of connections, such as bids submitted in the same tender. Ultimately, message passing in GNNs enables features to flow between any two nodes (e.g., bids) that are not (or appear not to be) directly connected, as long as there are common neighbors $N$ steps away connected through any combination of edges when the network has $N$ aggregation layers. 
For example, GNNs have shown promise in detecting inflated references in academic works by modeling them as a citation graph, capturing coordinated or artificial selection of references \citep{liu2022deep}. Message passing in GNNs is also important for collusion detection, potentially leading to better results when participants exhibit conditional behavior based on the general strategy of the group \citep{wu2020comprehensive}. 
Moreover, by not treating categorical variables as sparse matrices, GNNs reduce the number of features, resulting in less overfitting compared, for example, to feedforward NNs. 
GNN models that account for multiple relations are often called relational graph neural networks (R-GNNs) or relational graph convolutional networks (R-GCNs).

In this paper, we first develop models based on NN and GNN for collusion prediction in single markets (Phase I). GNNs explicitly allow the incorporation of networks and network structures, which are prevalent in collusion and key to improving predictive performance. For each market, we compare the predictive performance of GNNs to that of NNs. Second, we propose an out-of-distribution generalization approach (Phase II) in which the models estimated from a single market are used to detect collusive bids in new markets for which no training data is available. This approach addresses a common limitation of machine learning models, which often struggle when training data is missing.

We  contribute to the field of collusion detection by developing advanced detection methods, namely GNNs, which can outperform NNs and other machine learning techniques, strengthening efforts to maintain market integrity and prevent anti-competitive behavior worldwide. Our study is one of the first to use GNNs for economic applications and shows their potential, particularly when networks are prevalent. 

The paper is structured as follows: Section \ref{LiteratureReview} provides an overview of the existing literature. Section \ref{GNN} introduces the GNN notation and key concepts related to message passing, normalization, and regularization. Section \ref{ResearchMethodology} describes the empirical application, including the datasets and the implementation of the methods. 
Sections \ref{Results I} and \ref{Results II} present and discuss the empirical results, focusing first on prediction in a single market (\textit{Results I}) and then on the out-of-distribution generalization of results (\textit{Results II}). Section \ref{finalremarks} concludes with final remarks.
    
    \section{Literature Review\label{LiteratureReview}}

    The ABA is an auction system that determines the winning bid based on bid averages within a tender. The average bid price $A1$ can be defined as:
    
    \begin{equation}
    A1 = \frac{\sum_{i=1}^{n} \text{bid}_i}{n},
    \end{equation}
    
    where $n$ is the number of participants in the auction and can be trimmed to exclude extreme values on both sides. A second average, $A2$, is calculated as the average of bids lower than $A1$. The specific rules for selecting the winning bid using ABA may vary among tenders, with the winning bid usually being the lowest bid above $A2$ \citep{conley2016detecting}.
    
    ABA analysis offers a method for collusion detection by comparing how much a given group can influence the average bid price in a series of auctions. This is done by regressing previously defined subgroups over the average price A1, allowing a threshold to be defined to identify groups strongly associated with manipulating the average price \citep{conley2016detecting}. 
    
    When the average price $A1$ increases systematically and significantly for a given group, it may indicate the presence of collusion among bidders. However, this approach has limitations because it may not be sensitive enough to detect more subtle forms of collusion or distinguish between legitimate price increases and those resulting from collusion \citep{bajari2002detecting}. Moreover, to identify a suitable threshold, it requires groups to be defined before applying the regression model, which can be challenging if the groups are not identified accurately. 
    
    Theory-driven models for collusion detection rely on established frameworks and prior knowledge about the characteristics and mechanisms of collusion. These models typically involve the use of screening variables derived from the original input data through transformations designed to capture key aspects of collusive behavior \citep{jacquemin1989cartels}. Examples of such screening variables include measures of bid dispersion or asymmetries in bid distributions \citep{wallimann2022machine}.
    
    By incorporating domain knowledge into the modeling process, theory-driven models have the potential to improve the performance of collusion detection algorithms by focusing on features more likely to be associated with collusive behavior. Recent studies have explored the use of various machine learning techniques, such as random forests for classifying incomplete cartels \citep{wallimann2022machine} and support vector machines (SVMs) for classifying collusive bids \citep{rodriguez2022collusion} in combination with screening variables. These studies have reported improvements in model performance compared to purely data-driven approaches.
    
    State-of-the-art models for collusion detection usually involve various machine learning and statistical techniques. These models rely on features extracted from the data to identify patterns in bid prices, bid allocation, and other variables that may indicate collusion. Examples of relatively modern approaches include linear models such as stochastic gradient descent (SGD); ensemble methods such as extremely randomized trees (extra trees), random forest, AdaBoost, and gradient boosting; support vector machines (SVMs) such as C-support vector classification (C-SVC); nearest neighbors models such as k-nearest neighbors (k-NN); and neural network models such as multi-layer perceptron (MLP) \citep{rodriguez2022collusion,wallimann2022machine,imhof2021detecting}. Although these models have shown promise in detecting collusion, their performance can be limited by factors such as data quality, feature selection, and the inherent complexities of the collusion phenomenon.

    Various machine learning techniques have been applied to the problem of collusion detection with varying degrees of success. For example, \citet{rodriguez2022collusion} compared the performance of 11 machine learning models (SGD, extra trees, random forest, AdaBoost, gradient boosting, SVM, k-NN, MLP, Bernoulli Naive Bayes, Gaussian Naive Bayes, and Gaussian process) for classifying a bid as either collusive or non-collusive. They found that, in terms of balanced accuracy, ensemble methods (extra trees, random forest, AdaBoost, and gradient boosting) performed best in many cases, but, in most cases, the other models did not.
    These mixed results suggest that the optimal machine learning technique for collusion detection may be context-dependent and influenced by factors such as data quality, feature selection and engineering, and model complexity. Additionally, the use of convolution layers on graph structures, such as image structures, has been shown to be effective in improving performance for collusion detection and uncovering complex relationship patterns in the data, especially when sufficient resources are available for model training \citep{huber2023flagging}.

    A common challenge in using machine learning techniques for collusion detection is the imbalance in classification labels: the number of non-collusive instances substantially outweighs the number of collusive instances \citep{rodriguez2022collusion}. This imbalance can lead to high rates of false negatives because models may be biased towards classifying instances as non-collusive. To address this problem, researchers have employed various techniques, such as oversampling, undersampling, or using synthetic data to balance the training dataset \citep{chawla2002smote}. However, these approaches are not always effective at improving the detection of collusive instances, and there remains room for improvement in addressing class imbalance in collusion contexts.
    
      GNNs are a class of neural network models specifically designed to operate on graph-structured data, which is characterized by entities and their relationships \citep{wu2020comprehensive}. With their ability to learn latent representations (non-observable transformations of the input) of entities and their relationships, GNNs can potentially improve the performance of collusion detection algorithms by identifying subtle patterns, such as groups of entities engaging in collusive behavior that may be difficult to detect using traditional machine learning techniques.
    
    Unlike many traditional machine learning models, GNNs do not assume independence among observations, which is particularly relevant for collusion detection given that bidders' behaviors may be conditionally dependent on each other \citep{wu2020comprehensive}. By modeling the relationships between entities explicitly, GNNs can account for the complex interactions and dependencies that may exist among bidders and thus better capture the underlying structure of collusive behavior \citep{wu2020comprehensive}. This ability to model dependencies can potentially lead to better results in detecting collusion, especially in cases in which participants exhibit conditional behavior based on the overall group strategy. Moreover, by considering the relationships and dependencies among bidders, GNNs have the potential to outperform other machine learning techniques in cases in which participants' behavior depends on the group's overall strategy. In fact, GNNs are often the standard approach for many collusion detection tasks, even though they generally require a higher number of labeled observations \citep{zhou2023fraudauditor}.

    GNNs are useful in contexts in which Euclidean spaces are insufficient to represent the relationship between variables. Specific GNN models include convolutional GNN, recurrent GNN, graph auto-encoders, spatial-temporal GNN, and relational graph convolutional networks \citep{wu2020comprehensive,nicolicioiu2019recurrent,wang2017mgae}. Tasks usually involve the classification of nodes (also referred to as entities), links, or the graphs themselves \citep{wu2020comprehensive}. 
    
    Convolutional GNNs are GNNs capable of transforming and aggregating features from neighboring nodes to the respective nodes. This process, which is often referred to as message passing, involved propagating features from other observations in the dataset to the connected neighbors. The propagation can extend as far as the number of layers from the source node. In turn, relational graph convolutional networks are GNNs that not only propagate message passing but also allow for different types of edges \citep{schlichtkrull2018modeling}. This means that the same feature, from both the node itself and from its neighbors, can be accounted for multiple times with different weights assigned to each relation type. 

    In conclusion, the use of GNNs in collusion detection represents a promising avenue for research because they have the potential to overcome some of the limitations of traditional machine learning techniques. GNNs can explicitly or implicitly model edges (connections) and node representations without assuming independence among observations, thus capturing conditional behavior \citep{chien2020adaptive,wu2020comprehensive}.

    \section{Introduction to Graph Neural Networks \label{GNN}}
    
         We use relational graph convolutional networks (R-GCNs) because of their ability to assign different weights to different types of edges. This allows us to connect a bid, for example, to another bid made in the same tender, as well as to the previous and next bids from the same company while assigning them different weights without changing the original features from the source node. Additionally, the convolutional layers allow the relatively large feature space inherent in GNNs to be reduced, facilitating convergence and stability of the classification. 
    
    The most important task of GNNs is to find the best node representation for a given function. A graph $G$ with $\mu \in \{1,..,m\}$ nodes and feature dimension $\hat{f} \in \{1,..,f\}$ can have its raw input feature vectors $\beta_{\mu}$ represented as a matrix $X$ of features $x_{\mu, \hat{f}}$ with dimensions $m \times f$:
    
    \begin{equation}
    \label{eq:matrix1}
    X = \begin{bmatrix} 
    x_{1,1} &  .. & x_{1,f}  \\
    .. &  .. & ..            \\
    x_{m,1} &  .. & x_{m,f}  \\
    \end{bmatrix}
    = \begin{bmatrix} 
    \beta_1  \\
    ..            \\
    \beta_m  \\
    \end{bmatrix}
    \end{equation}

    where each $\mu \in \{1,..,m\}$ represents an individual node, i.e., a bid. $\beta$ is defined as $\beta_{\mu} = (x_{\mu, 1},..,x_{\mu, f}) \forall \mu$ , where $x_{\mu, \hat{f}}$ represents the  $\hat{f}$-th feature of the $\mu$-th node. Let $h_{\mu}^{(0)}$ be the raw representation of the input row (node) $\mu$ on time stamp $k=0$, such that 

    \begin{equation}
    \label{eq:matrix2}
    h_{\mu}^{(0)} = \begin{bmatrix} 
    x_{\mu,1} &  .. & x_{\mu,f}  \\
    \end{bmatrix} = \beta_{\mu} \quad \forall \mu \in \{1,..,m\}
    \end{equation}

    represents input features. The number of layers in the GNN is represented as  $k \in \{0,..,K-1\}$, where $k=0$ is the raw input, $k=1$ is the first layer, and so on. In the GNN context, the output of each layer will include a combination of its neighbors. This aggregation step can be performed, for example, as a sum over the neighbors, including the source node features such that the propagation continues recursively for any node $\mu$ with $v \in \eta(\mu)$ neighbors at layer $k$:

    \begin{equation}
    \label{MP-1}
    h_\mu^{(k+1)} =  h_\mu^{(k)}  +  \sum_{v \epsilon \eta(\mu)} h_v^{(k)} \quad \forall k \in \{0,..,K-1\} \text{ and for } \forall \mu \in \{1,..,m\}.
    \end{equation}
    
\begin{figure}[H] %[!ht]
        \centering
        \includegraphics[scale=0.8]{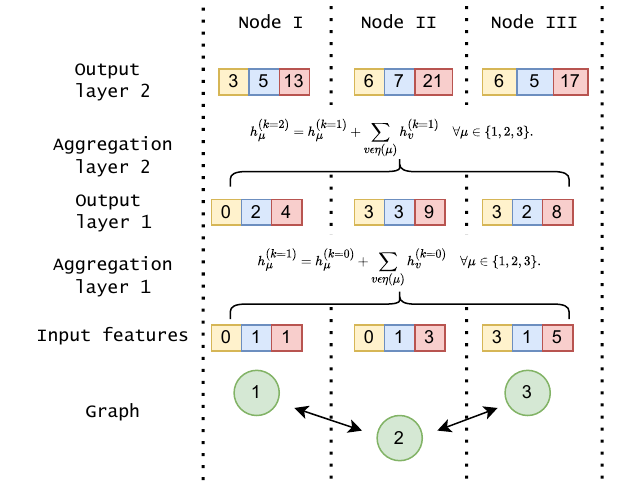}
        \caption{Message passing example for three nodes using the simplified Equation \eqref{MP-1}. Nodes I and III are connected to node II, and node II is connected to nodes I and III. Self-connections are present for all nodes.}
        \label{fig:mp-example}
    \end{figure}

    The simplified aggregation function presented in Equation \eqref{MP-1} is also referred to as message passing. Figure \ref{fig:mp-example} shows an example of message passing for two aggregation layers.
    
    Figure \ref{fig:mp-example} demonstrates the effects of feature smoothing and feature propagation. For example, $x_{3,3} = 5$ is smoothed at the output to a value closer to the average of the group (13, 21, 17) and propagated to nodes I and II even though nodes I and III are not directly connected. This is an important feature for collusion detection because known labels could be associated with specific features that would cause the model to fit to the specific nodes rather than the group. Analogously, message passing makes it possible to detect coordination because the model receives the features aggregated from a group instead of treating single observations in isolation. 
    
    The standard approach to most GNN tasks involves applying weights to the features, followed by an activation function. By adding weight vectors $w$ to the equation, the output representation becomes:
    
    \begin{equation}  h_\mu^{(k+1)} =  h_\mu^{(k)}w_\mu^{(k+1)}  +  \sum_{v \epsilon \eta(\mu)} h_v^{(k)}w_v^{(k+1)} \quad \forall k, \forall \mu,
    \end{equation}
    
    where $w_{\mu}$ and $w_{v}$ are the weights to be optimized. The size of the weight vector $w$ dictates the number of output features of the respective hidden layer, thus allowing for convolution if the size is smaller than $f$. To account for non-linearity, an activation function $\sigma$, such as rectified linear unit (ReLU) or sigmoid functions, can be applied to the last layer \citep{pmlr-v97-wu19e}:
    
    \begin{equation}
    \label{eq:GNNfinallinear}
     h_{\mu}^{(k+1)} = \sigma\left(h_{\mu}^{(k)}w_{\mu}^{(k+1)}+    \sum_{v \epsilon \eta(\mu)} h_v^{(k)}w_v^{(k+1)}  \right) \quad \forall k, \forall \mu.
    \end{equation}

    An alternative way to represent this structure for all nodes simultaneously is by using the adjacency matrix $\hat{A}$ to perform matrix multiplication with vectors $h_\mu^{(k)}$ and $h_v^{(k)}$. The adjacency matrix $\hat{A}$ for this graph $G$, excluding the value of the nodes 
    themselves (i.e., $\hat{a}_{ii}=0$), becomes a $i\times j$ square matrix with dimension $m \times m$, where nodes $i$ connect to nodes $j$. The 
    values $\hat{a}_{ij}$ are equal to 1 if the connection exists and are equal to 0 otherwise. By adding the identity matrix $I$ to $\hat{A}$, we connect the node's own value to the following function:
    
    \begin{equation} \overline{A} = \hat{A} + I = \begin{bmatrix}
    0 & .. & \hat{a}_{1,m}  \\
    .. & 0 & ..  \\
    \hat{a}_{m,1} & .. & 0 \\
    \end{bmatrix} + \begin{bmatrix}
    1 & .. & 0 \\
    .. & 1 & ..  \\
    0 & .. & 1  \\
    \end{bmatrix} 
    = \begin{bmatrix}
    1 & .. & \hat{a}_{1,m}  \\
    .. & 1 & ..  \\
    \hat{a}_{m,1} & .. & 1 \\
    \end{bmatrix}.
    \end{equation}
    
    If $H^{(k)}$ is the vectorial representation of $h_{\mu}^{(k)}$ $\forall \mu$ at time $k$, and if $W^{(k+1)}$ is the equivalent representation of the weights $w^{(k+1)}$ at time $k+1$, then $H^{(k)}$ and $W^{(k+1)}$ can be defined as:

    \begin{equation}
    H^{(k)} = \begin{bmatrix}
         h_{1}^{(k)} \\
         .. \\
          h_{m}^{(k)}\\
    \end{bmatrix} \quad \forall k , \quad
    W^{(k+1)} = \begin{bmatrix}
         w_{1}^{(k+1)} \\
         .. \\
          w_{m}^{(k+1)}\\
    \end{bmatrix}  \quad \forall k
    \end{equation} 
    
    such that Equation \eqref{eq:GNNfinallinear} can be written in terms of $\overline{A}$ and $W^{(k+1)}$ \citep{pmlr-v97-wu19e}:

    \begin{equation}
    \label{endGNN}
    H^{(k+1)} = \sigma \left(\overline{A}H^{(k)}W^{(k+1)}\right) \quad \forall k
    \end{equation}
    
    where $H^{(k+1)} = \hat{Y}$ if $k=K-1$, i.e., the output of the training process at the last layer with an activation function $\sigma$.

\subsection{Regularizing the Graph Structure}

    To account for the effect that nodes with more neighbors would inherently have higher values, the Degree Matrix $D$ can be applied to Equation \eqref{endGNN}. The values $d_{ij}$ of $D$ are equal to 0, except for the diagonal. The diagonal is built upon the values of $\overline{A}$ and is defined as:
    
    \begin{equation} 
    d_{ii} = \sum_{j=1}^m \overline{a}_{ij} \quad \forall i \in \{1,..,m\}.
    \end{equation}
    
    The inverse $D^{-1}$ is a matrix of values between 0 and 1, penalizing nodes with  more neighbors. Applying $D^{-1}$ to Equation \eqref{endGNN} results in the following equation \citep{kipf2016semi}:
    
    \begin{equation}
    \label{GNNNorml}
    H^{(k+1)} = \sigma (D^{-1}\overline{A}H^{(k)}W^{(k+1)}) \quad \forall k.
    \end{equation}
    
    An alternative approach to $D^{-1}$ is to apply $D^{-1/2}$ to both sides of $\overline{A}$ to account for neighbor values once for the lines (connections) and once for the columns (being connected to):
        
    \begin{equation} \tilde{A} = D^{(-1/2)}\overline{A}D^{(-1/2)}.
    \end{equation}
    
    Hence, the main equation for updating the node representations becomes \citep{kipf2016semi}:
    
    \begin{equation}
    \label{endCGNNmg}
    H^{(k+1)} = \sigma \left(\tilde{A}H^{(k)}W^{(k+1)}\right) \quad \forall k.
    \end{equation}
    
    Equation \eqref{endCGNNmg} is the base notation we will use in this work and is adjusted to account for multiple types of edges. 

\subsection{Relational Graph Convolutional Networks}
    
\begin{figure}[!h]
        \centering
        \includegraphics[scale=0.28]{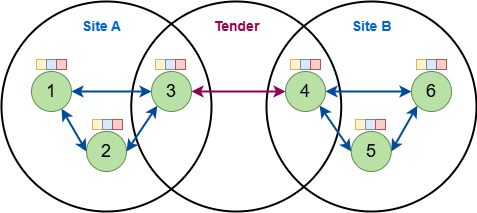}
        \caption{Different nodes as bids connected by different relationships (company site vs. tender) simultaneously.}
        \label{fig:RGNN-1}
    \end{figure}

    Relational graph convolutional networks (R-GCNs), an extension of graph convolutional networks (GCNs), are designed to handle graphs with multiple types of relations between nodes. These networks have been used for link prediction and entity classification tasks in knowledge graphs but can also be applied to other domains, such as collusion detection. Figure \ref{fig:RGNN-1} provides an example of R-GCNs.

    The fundamental idea behind R-GCNs is to incorporate relation-specific transformations in addition to node features. In the context of collusion detection, different relations can be used to account for information sharing from nodes connected to the previous and next bids of the same company, or for companies participating in the same tender or in the same region. When these different types of relations are considered, R-GCNs can be expected to be more effective in capturing patterns indicative of collusion behavior.
    
    The main equation for updating node representations in an R-GCN is given by \citep{schlichtkrull2018modeling}:
    
    \begin{equation}
    \label{}
     H^{(k+1)} = \sigma \left( \sum_{r=1}^{R} \tilde{A}^{(r)}H^{(k)}W^{(k+1,r)} \right) \quad \forall k
    \end{equation}
    
    where $\tilde{A}^{(r)}$ is the normalized adjacency matrix for each edge type $r \in \{1,..,R\}$,  $W^{(k+1,r)}$ is the weight matrix for relation $r \in \{1,..,R\}$ at layer $k+1$ for $k \in \{0,..,K-1\}$. The normalization of the adjacency matrix is performed similarly to R-GCNs by having one adjacency matrix for each edge type:
    
    \begin{equation}
    \label{}
    \tilde{A}^{(r)} = D^{(r)(-1/2)}\overline{A}^{(r)}D^{(r)(-1/2)} \quad \forall r \in \{1,..,R\}
    \end{equation}
    
    where $D^{(r)}$ and $\overline{A}^{(r)}$ are specific to each $r$. Analogously, each edge type receives its own weight matrix $W^{(k+1,r)}$. By using R-GCNs, different types of relationships can be combined to generate richer node embeddings, making them more effective at detecting collusive behavior among companies participating in bids and tenders. The network can learn the importance of each type of relation and take advantage of this information to improve its predictions. This approach allows for the detection of complex patterns and interactions that might be missed when considering only a single type of relation, leading to a more accurate and robust collusion detection approach. 

    R-GCNs use the concepts of base and diagonal decomposition to simplify complexity and enhance efficiency, especially when dealing with graphs containing multiple types of relations. Base decomposition in R-GCNs is an approach designed to reduce the number of parameters, particularly when the number of relations $ R $ is high. This technique involves decomposing the relation-specific weight matrices into a shared base weight matrix and several smaller relation-specific diagonal matrices. Formally, the weight matrix for each relation type $ W^{(k+1,r)} $ can be decomposed as follows:
    
    \begin{equation}
    W^{(k+1,r)} = U^{(k+1)}V^{(k+1,r)}
    \end{equation}
    
    Here, $ U^{(k+1)} $ is the shared base weight matrix for layer $ k+1 $, and $ V^{(k+1,r)} $ is a diagonal matrix specific to relation type $ r $. This decomposition substantially reduces the number of parameters because $ U^{(k+1)} $ is shared across all relations and $ V^{(k+1,r)} $ contains far fewer parameters than the original $ W^{(k+1,r)} $.

    Diagonal decomposition further simplifies the model by assuming that the relation-specific weight matrices are diagonal. Under this scheme, the weight matrix for each relation type becomes a diagonal matrix, meaning that each feature only interacts with itself, not with other features. The diagonal weight matrix for relation type $r$ at layer $k+1$ is represented as:

    \begin{equation}
    W^{(k+1,r)} = diag(w^{(k+1,r)}) 
    \end{equation}

    where $diag(w^{(k+1,r)})$ is a diagonal matrix with the elements of the vector $w^{(k+1,r)}$ on its diagonal. This decomposition drastically reduces the complexity and the number of parameters in the network, making it computationally more efficient, albeit at the potential cost of expressive power.
    
    Both base and diagonal decompositions are strategies employed to balance the trade-off between the complexity of a model and its expressive capability. In scenarios with a large number of relation types, these decompositions can substantially improve the scalability and efficiency of R-GCNs. They are particularly useful in situations in which computational resources are limited or when the network needs to be highly responsive. By carefully choosing between base and diagonal decomposition, one can design R-GCNs that are both efficient and effective for a wide range of applications, including the detection of collusive behavior in bidding networks.

\subsection{R-GCNs for Collusion Detection}

    GNNs offer a range of architectures, each tailored to specific types of graph-related tasks. Among the notable architectures are graph convolutional networks (GCNs), which are well-suited to tasks like node classification and link prediction through their leveraging of local graph structures \citep{kipf2016semi}. GCNs excel at learning node representations through neighbor feature aggregation, allowing them to capture essential connectivity patterns. Graph attention networks (GATs) introduce an attention mechanism, allowing for dynamic weighting of neighbor contributions, making them ideal for scenarios in which neighbor importance varies substantially \citep{velickovic_gat}. GraphSAGE facilitates inductive learning on graphs, enabling the model to generalize to unseen nodes by sampling and aggregating features from a fixed-size neighborhood, thus efficiently managing large graphs \citep{hamilton_graphsage}. Graph isomorphism networks (GINs) stand out for their ability to distinguish between different graph structures, closely mirroring the Weisfeiler-Lehman graph isomorphism test. GINs are particularly effective at tasks requiring a deep understanding of graph topology \citep{xu_gin}.
    
    Our decision to use R-GCNs to detect collusion among bids and tenders is based on the unique requirements of our application domain. R-GCNs are adept at processing graphs with multiple types of relationships, which are a common characteristic of collusion detection scenarios. In such scenarios, it is crucial to model various interactions, such as bids in the same tender or sequential bids from the same company, with distinct weights to identify patterns indicative of collusion. This makes it possible to detect complex collusion patterns that might be missed by architectures that either treat all edges equally or cannot differentiate between edge types. Additionally, comparing R-GCNs with NNs is facilitated by the capacity of R-GCNs to handle relational data, similar to how NNs process structured, non-relational data. This parallel enables a clear evaluation of the benefits and limitations inherent in applying deep learning methods to relational data, offering a fair assessment of the effectiveness of the basic architecture in detecting collusion.
    
    \section{Empirical Analysis\label{ResearchMethodology}}
    
\begin{figure}[!h]
    \centering
    \includegraphics[scale=0.4]{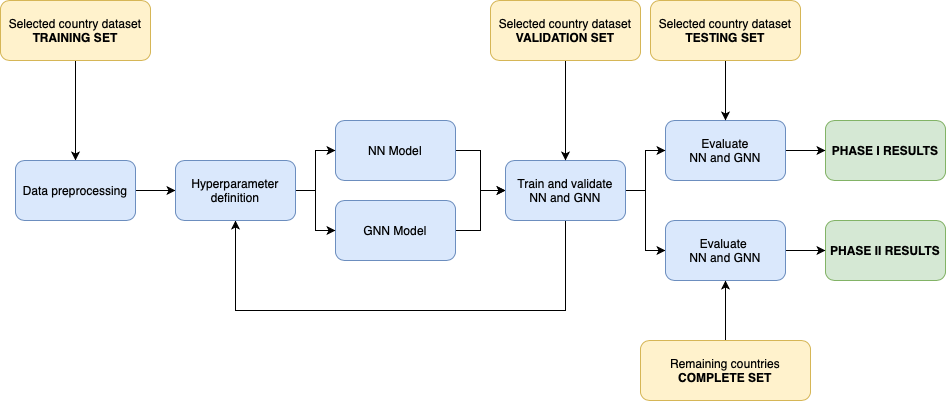}
    \caption{Method Overview}
    \label{fig:method-overview}
\end{figure}

        Our study compares the performance of GNNs and NNs for collusion detection when trained and evaluated on the bid-rigging datasets\footnote{The raw data are tabular. An extract of the raw data from Japan can be found in Appendix \ref{Appendix}.} from \cite{rodriguez2022collusion}. It also measures the performance of the models when applied to data from other countries during inference. We train one GNN and one NN model for each of the following countries or regions:\footnote{For the sake of simplicity, we refer to the datasets of the two Swiss regions in the remainder of this paper as country datasets.} Japan, Brazil, US, Italy, Ticino (Switzerland) and St. Gallen and Graubünden (Switzerland). Subsequently, we evaluate the performance of each model on the testing set of the same country (Phase I results) and then on the complete dataset of each of the remaining countries (Phase II results). Figure \ref{fig:method-overview} gives an overview of the procedure. 

In Phase I, we prepare the data, select the features to be used, and train the models. During training on each dataset, a three-fold cross-validation grid search is applied, consisting of eight different hyperparameter configurations. The best performing configuration is selected to evaluate the model's performance on the test set. This process of re-sampling, cross-validation grid search, training, and testing is repeated 10 times to compute the average and standard deviation of the results. The parameters tested and the number of re-sampling runs are defined based on computing resources and model stability. In Phase II, we apply the models trained on one country's dataset in Phase I to the remaining datasets and evaluate performance.

\subsection{Datasets and Data Sources\label{Dataset}}

     The primary dataset used in this study is derived from the work of \citet{rodriguez2022collusion},\footnote{Dataset available at: https://www.sciencedirect.com/science/article/pii/S0926580521004982\#ec0010} encompassing procurement data from Japan \citep{ishii2014bid}, Brazil \citep{signor2019not}, the US \citep{porter1999ohio}, two regions in Switzerland \citep{imhof2017simple,imhof2018screening}, and Italy \citep{conley2016detecting}. 
    
    The Brazilian data pertains to bid rigging in Petrobras's infrastructure projects, primarily from 2002 to 2013, involving a cartel known as the ``Club of 16". The Italian dataset covers road construction auctions in Turin from 2000 to 2003, with a focus on bid rigging by numerous firms. In Japan, the dataset includes building construction and civil engineering contracts in Okinawa from 2003 to 2007, highlighting a collusion-facilitated environment. The two Swiss datasets, one from the Canton of Ticino and the other from the cantons of St. Gallen and Graubünden, range from the mid-1990s to 2010 and detail the behavior of severe bid-rigging cartels in the road construction and civil engineering sectors. Finally, the US dataset provides a unique perspective from the dairy sector, focusing on school milk procurement in Ohio between 1980 and 1990, with evidence of substantial bid rigging leading to increased prices. 
    
    In addition to the geographical and sectoral diversity of the datasets, the auction methods employed in these cases also vary and play an important role in the dynamics of bid rigging. Brazil and Japan predominantly used standard open bidding processes, in which bids are publicly submitted and the lowest bid typically wins. Italy, however, employed the ABA method, wherein the contract is awarded to the bid closest to a trimmed average of all bids, a method that can inadvertently incentivize bid coordination and manipulation. The US dataset primarily uses a straightforward lowest-bid-wins approach, standard in many public procurement processes. In the Swiss datasets, the lowest bid also typically wins.

\subsection{Features \label{NNFeatures}}

    The variables from the dataset \citep{rodriguez2022collusion} which are used as features for both training and later evaluation can be seen in Table \ref{table:nnfeatures}.
    
    \begin{table}[!ht]
        \captionsetup{justification=centering}
        \caption{Features for NN and GNN}
        \label{table:nnfeatures}
        \begin{tabular}{cl}
        \hline
        \textbf{Feature} & \textbf{Description} \\
        \hline
        $Bid\_value$ & Price offered by the company \\ %Value bid by the company \\
        $Number\_bids$ & Total number of bids in a given tender \\ %Total number of bids of a given company among all Tenders \\
        $Winner$ &Binary variable: 1 if bid is winner of the tender, 0 otherwise \\ %Binary variable if the bid was successful \\
        $CV$ & Coefficient of variation\\
        $SPD$ & Spread inside tender \\
        $DIFFP$ & Difference between the two lowest bids within a tender \\ %Difference between the highest and lowest bid within a Tender \\
        $RD$ & Relative distance \\ %Relative Distance between bids \\
        $KURT$ & Excess kurtosis \\
        $SKEW$ & Skewness \\
        $KSTEST$ & Kolmogorov-Smirnov test \\
        $PTE$* & Pre-tender Estimate (PTE) for a given tender \\
        $Date$* & Date of the tender \\
        \textit{Difference Bid/PTE}* & Distance between bid value and PTE \\
        \hline
        \end{tabular}
        \footnotesize *Features not included during training.
    \end{table}

     All values are normalized between 0 and 1. Variables \textit{CV, SPD, DIFFP, RD, KURT, SKEW,} and \textit{KSTEST} are screening variables from the original dataset. Details on how the screening variables are calculated can be found in Appendix \ref{Appendix-b}.

\begin{figure}[H] %[!ht]
      \centering
      \makebox[\textwidth][c]{%
        \includegraphics[width=1.0\textwidth]{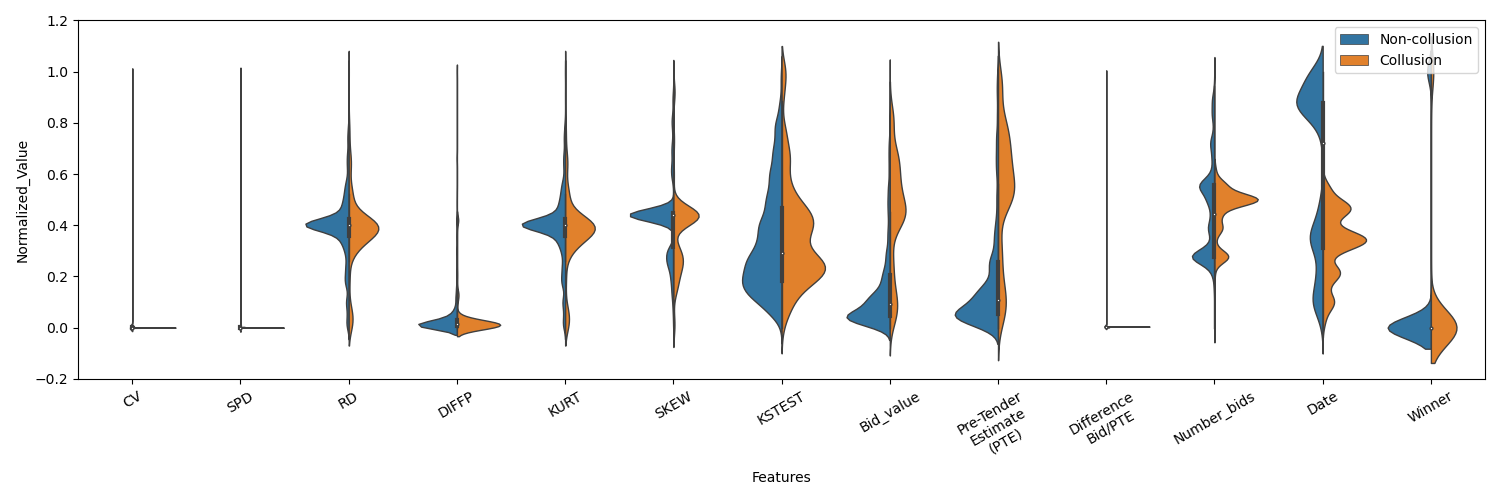}
      }
      \caption{Distribution of variables among collusive and non-collusive bids in the Japanese dataset}
      \label{fig:variables_comparison}
    \end{figure}

    We exclude $Date$ as a feature during training because, for example, in the Japanese dataset, there is evidence of bid-rigging only at the beginning of the recorded time period, as shown in Figure \ref{fig:variables_comparison}. Furthermore, we decide to eliminate the variables \textit{Difference Bid/PTE} and \textit{PTE} because they are not included in all datasets. Hence, the features used across the results section are as follows: \textit{Bid\_value, Number\_bids, Winner, CV, SPD, DIFFP, RD, KURT, SKEW}, and \textit{KSTEST}.

    Figure \ref{fig:variables_comparison} illustrates the distribution of features in the two classes of the Japanese dataset to identify whether and how the distributions differ between non-collusive and collusive bids. The distributions of \textit{Date} show that collusion took place at the beginning of the recorded time period. Additionally, the respective distributions of \textit{Bid\_value} and \textit{PTE} for collusive bids differ from those of non-collusive bids, indicating that collusion was more frequent in larger projects with higher PTE values. The distribution of the variables across the two classes in the other datasets is shown in Appendix \ref{Appendix-c}.
    
\subsection{GNN Features, Nodes, and Edges\label{GNNFeatures}}

    The features of the GNN model are the same as those of the NN model (as detailed in section \textit{\ref{NNFeatures} \nameref{NNFeatures}}) except that their values are influenced by the connected neighbors in the graph. Each bid is a row in the dataset and is therefore treated as a node in the graph. The complete list of edge types for the GNN model can be seen in Table \ref{table:edges_relations}.

    \begin{table}[!ht]
    \captionsetup{justification=centering} 
    \caption{GNN edges between nodes}
    \label{table:edges_relations}
    \resizebox{\textwidth}{!}{%
    \begin{tabular}{cc}
    \hline
    \textbf{Edges relations} & \textbf{Description}                                                             \\ \hline
    Tender                   & Tender ID, connect bids made in the same tender                \\
    Competitor               & Company ID, connect bids made by the same company               \\
    Location                 & Tender Geographical Location ID, connect bids with the same Tender Location ID \\ 
    Site                     & Company Geographical Site ID, connect bids with the same Company Site ID \\ \hline
    \end{tabular}%
    }
    \end{table}

    The edges available for each dataset are shown in Table \ref{tab:gnn_edges_all_models}. Japan, Brazil, and Italy have three of the four edge relations in their datasets. In the St. Gallen and Graubünden, Ticino, and US datasets, only one or two edge types are available. Therefore, we expect that the performance of the GNN trained with Swiss or US data will be lower than that of the GNN trained using all three available edges, which together potentially contain more relevant information. 

     \begin{table}[!ht]
         \caption{Edges available to each dataset}
         \centering
        \begin{tabular}{lcccc}
        \hline
        \textbf{Dataset} & \textbf{Tender ID} & \textbf{Company ID} & \textbf{Location ID}  & \textbf{Site ID}  \\
        \hline
        Japan & $\boxtimes$ & $\boxtimes$ & $\boxtimes$ & $\square$ \\
        Brazil & $\boxtimes$ & $\boxtimes$ & $\boxtimes$ & $\square$\\
        Italy & $\boxtimes$ & $\boxtimes$ & $\square$ & $\boxtimes$\\
        United States (US) & $\boxtimes$ & $\boxtimes$ & $\square$ & $\square$  \\
        St. Gallen and Graubünden & $\boxtimes$ & $\square$ & $\square$ & $\square$\\
        Ticino & $\boxtimes$ & $\square$ & $\square$ & $\square$\\
        \hline
        \end{tabular}
        \label{tab:gnn_edges_all_models}
    \end{table}
    
    To build the GNN, the node connections must be defined in advance. Once the structure is defined, information will flow between the features of nodes that share a connection if the corresponding edge variable is present. We connect all nodes to all other nodes for which the relation matches, except the \textit{Company ID} variable, for which we connect only the previous and the next bid from the same company for any given bid instead of connecting all bids from the same company. This approach introduces a temporal element to the network and reduces excessive information flow, thus lowering the risk of excessive feature smoothing. Because the Italian data lacks date information, each bid is connected to the previous and next bid from the same company based on the order in the raw data.
    
\begin{figure}[H]
      \centering
      \includegraphics[width=0.75\textwidth]{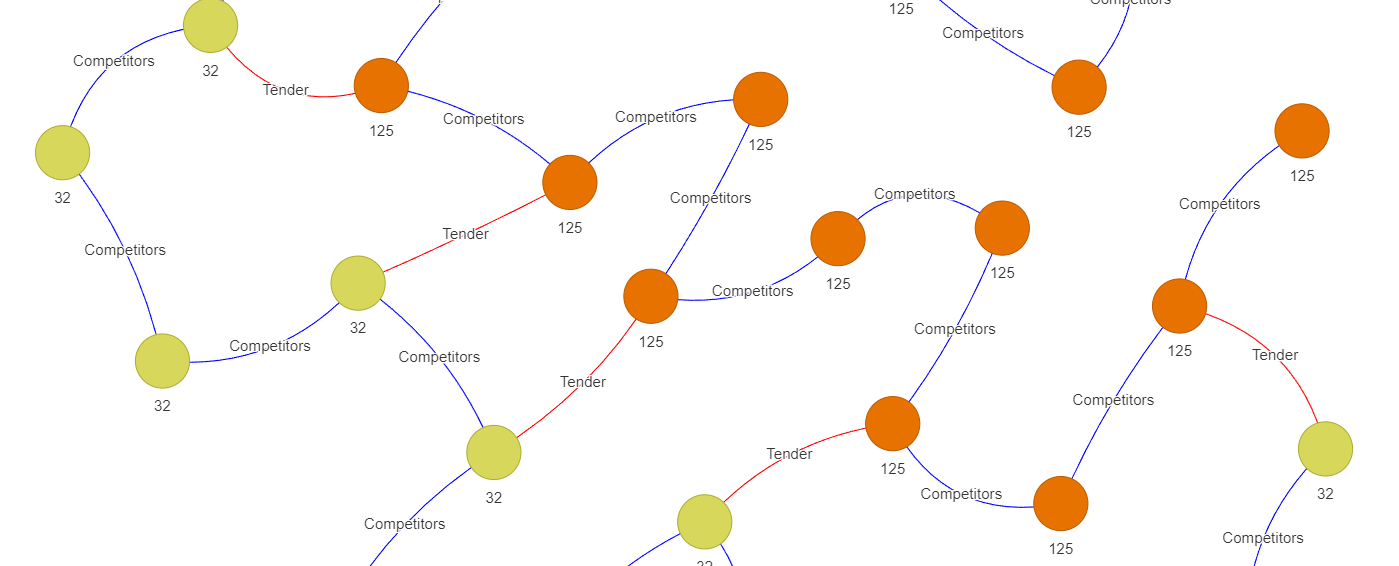}
      \caption{Different nodes as bids connected by different relationships (competitor vs. tender) simultaneously.}
      \label{fig:nodesexample}
    \end{figure}

     Figure \ref{fig:nodesexample} provides a comprehensive example visualization of the nodes present in the graph for two random companies, showing the two-neighbors-only Company ID connection.
    
    In this visualization, node features (such as bid values) are not shown, only their relations. Nodes of the same color represent different bids from the same company. The lines connecting each node are the edges of the graph and represent their relations. 
    
    In this example, the blue edges represent connections between bids from the same company, whereas the red edges represent connections between bids made within the same tender. For the company edge, we connect only the previous bid to the next bid, whereas for the tender edge, we connect all bids within the same tender (as well as for all other edges except for \textit{Company ID}). This configuration also allows information to flow between companies that collaborate over time without depending explicitly on the \textit{Date} variable because bids from the same company are connected in chronological order.
    
     Every connection in the graph is bi-directional, meaning that each edge is duplicated in the opposite direction. This characteristic inherently increases the number of edges in the graph, making it considerably larger than the count of nodes. This bidirectional edge representation allows for more precise modeling of relationships among the nodes, accommodating the potential for interactions to flow both ways and capturing the reciprocal nature of these relationships. 
    
     Because information flows both ways along each edge, using more than one layer causes the information to bounce back, being aggregated and transformed with each layer further away from the source nodes. Each layer passes information from nodes one connection away from the source node, so a three-layer model would propagate the message to nodes three layers away for any feature in the connected graph. This makes the R-GCN a closed cycle, unlike directed acyclic graphs (DAGs), in which information flows in only one direction.

\subsection{Loss Function\label{Loss}}

    The target variable used to train the model is \textit{Collusive Competitor Original}, a binary variable from the original dataset \citep{rodriguez2022collusion}, with the value 1 indicating a collusive bid and 0 a non-collusive bid.
    
    To address the imbalance in the dataset,\footnote{A descriptive analysis of all datasets can be found in Appendix \ref{Appendix-d}.} we adjust the loss function to give more weight to the less frequent class (usually collusion). We achieve this by using the frequency of collusive bids, $n_{collusion}$, and the frequency of non-collusive bids, $n_{non-collusion}$, from the respective country dataset. As shown in Equation \eqref{equation:weights_coll} and Equation \eqref{equation:weights}, we calculate the weights $\theta_{collusion}$ and $\theta_{non-collusion}$ inversely proportional to their frequencies:
    
    \begin{equation}
    \label{equation:weights_coll}
        \theta_{collusion} = \frac{1}{n_{collusion}},
    \end{equation}

    and therefore

    \begin{equation}
    \label{equation:weights}
        \theta_{non-collusion} = \frac{1}{n_{non-collusion}},
    \end{equation}
        
    We modified the cross-entropy loss function, used as the loss function in our model, to account for the class imbalance in our datasets. Our weighted cross-entropy loss function is defined as:
    %This function combines softmax activation with a weighted cross-entropy loss, defined as:
    
    \begin{equation}
        L(y, \hat{y}) = -\theta_{\text{collusion}} \cdot y \log(\hat{y}) - \theta_{\text{non-collusion}} \cdot (1 - y) \log(1 - \hat{y}).
    \end{equation}
    
    In this equation, $L$ represents the loss, $y$ is the true label, and  $\hat{y} $ is the predicted probability for the positive class (collusion), which is the output of $H^{(k)}$ when $k = K$. The weights $\theta_{\text{collusion}}$ and $\theta_{\text{non-collusion}}$ are incorporated to address the class imbalance, with $\theta_{\text{collusion}}$ weighting the loss for the collusion class and $\theta_{\text{non-collusion}}$ weighting the loss for the non-collusion class.

\subsection{Data Splitting Strategy\label{DataSplitting}}

    For effective model training and validation, it is important to split the dataset strategically into training, testing, and validation sets. This split considers the unique companies and their respective bids. To avoid overfitting and to improve the generalization of the model, we labeled companies with at least one collusive bid as collusive for the purpose of dataset splitting, ensuring that the true classification labels of bids from the same company do not overlap between sets.
    
    The data is split as follows: 60\% of collusive and non-collusive companies are allocated to the training set, with the remaining 40\% split equally between testing and validation sets. Because the number of bids varies across companies, the actual percentage of bids in each dataset differs in every run but stabilizes over multiple iterations.

    The training and validation sets are used for a three-fold cross-validation grid search for hyperparameter tuning. The best-performing configuration is then selected to train the model using the combined training and validation sets. The F1 score validation set serves as the basis for the early stopping criterion, with a patience of 30 epochs. The performance of the model is evaluated on the test set using metrics such as the F1 score (as detailed in section \textit{\ref{evalmetrics} \nameref{evalmetrics}}).

\subsection{Model Architecture and Grid-search\label{Grid-search}}

    Both the NN and GNN models use a three-layer architecture. The first layer has an input-output dimension of (\textit{Input}, \textit{hidden-units}), the second layer (\textit{hidden-units, hidden-units/2}), and the final layer (\textit{hidden-units/2, Output}), where Input represents the number of input features and Output the number of classes (in our case, two). The first layer has a dropout rate of 20\%, the second layer 10\%, and the final layer 0\%. 

    The grid-search parameters used to determine the best hyperparameter configuration for the model are given in Table \ref{table:modelgridsearch}.
    
    \begin{table}[!ht]
    \centering
    \captionsetup{justification=centering} % Center the caption
    \caption{Grid-search parameters }
    \label{table:modelgridsearch}
    \begin{tabular}{cl}
    \hline
    \textbf{Parameters} & \multicolumn{1}{c}{\textbf{Values}}                                                                                                  \\ \hline

    learning rate              & \begin{tabular}[c]{@{}l@{}} [1e-2,1e-3] \end{tabular}                                       \\ \hline
    weight decay               & \begin{tabular}[c]{@{}l@{}}[1e-2,1e-3]\end{tabular}                                       \\ \hline
    hidden units               & \begin{tabular}[c]{@{}l@{}}[16,32]\end{tabular}                                       \\ \hline
    \end{tabular}
    \end{table}

    In Phase I, we train and evaluate each model separately for each country using that country's training, validation and test datasets. The process includes a three-fold cross-validation grid search to identify the best hyperparameter configuration, followed by training and evaluation of the model. This entire process is repeated over 10 runs for each of the two models per country. Finally, the mean and standard deviation of the evaluation metrics on the test set are computed. In Phase II, the procedure is similar but with a key difference: the two models developed in Phase I for a given country were applied separately to the complete dataset of each of the other countries (as presented in Figure \ref{fig:method-overview}). The following section defines the evaluation metrics used to analyze the results.

\subsection{Evaluation Metrics\label{evalmetrics}}

    This section presents the evaluation metrics used to assess the performance of our models: F1 score, balanced accuracy, precision, recall, receiver operating characteristic area under curve (ROC AUC), and precision-recall area under the curve (PR AUC). These metrics help us understand the trade-offs between precision and recall, as well as the overall robustness of the model. For this analysis, we consider collusive bids the positive class and non-collusive bids the negative class.
    
    The F1 score combines precision and recall into a single measure. Precision measures the accuracy of positive predictions (in our case, how often bids predicted as collusive are actually collusive). Recall, also known as sensitivity or the true positive rate (TPR), measures the proportion of actual positives (in our case, the true collusive bids) that are correctly identified by the model. Because the F1 score is defined as the harmonic mean of precision and recall, it balances the trade-off between these two measures, making it particularly useful in situations with imbalanced class distributions. By capturing this trade-off, the F1 score provides a balanced measure of a model's performance and is therefore an integral part of our evaluation process. The F1 score is defined as:
    
    \begin{equation}
    \text{F1 score} = 2 \cdot \frac{\text{Precision} \cdot \text{Recall}}{\text{Precision} + \text{Recall}}
     \end{equation}

    Precision and recall are calculated as follows:
    
    {\footnotesize
    \begin{equation}
    \text{Precision} = \frac{\text{True Positives}}{\text{True Positives} + \text{False Positives}}, \quad \text{Recall}=\text{TPR} = \frac{\text{True Positives}}{\text{True Positives} + \text{False Negatives}}.
    \end{equation}}

    Balanced accuracy is another important metric and is particularly useful in the context of imbalanced datasets. Unlike standard accuracy, which can be misleading when one class (usually non-collusive) dominates the predictions, balanced accuracy accounts for the performance of both classes equally. It measures the average accuracy obtained from both classes and is calculated as the arithmetic mean of recall and the true negative rate (TNR, also known as specificity):

    \begin{equation}
    \text{Balanced Accuracy} = \frac{1}{2} \cdot (\text{Recall} + \text{TNR})
    \end{equation}

    where \( TNR = \frac{\text{True Negatives}}{\text{True Negatives} + \text{False Positives}} \).

    We also evaluated precision and recall independently of each other. Precision is particularly important in the context of collusion detection because it measures the efficiency of our target prediction label regardless of the frequency of non-collusive bids in the dataset. In turn, assessing recall allows us to understand how well the model identifies true collusive bids. By considering recall and precision, we can determine whether the performance of the model is compromised, such as when it defaults to predicting the same class for all instances.

    ROC AUC is a performance measurement for classification problems at various threshold settings. It is defined as the area under the ROC curve, which plots the TPR against the false positive rate (FPR) at different thresholds. The ROC AUC score ranges from 0 to 1, with 1 representing a perfect model and 0.5 indicating no discriminative power. Mathematically, it is expressed as:
    
    \begin{equation}
    \text{ROC AUC} = \int_{0}^{1} TPR(FPR) \, dFPR
    \end{equation}
    
    where \( FPR = \frac{\text{False Positives}}{\text{False Positives} + \text{True Negatives}} \).
        
    PR AUC measures the trade-off between precision and recall for different threshold values, making it particularly useful for imbalanced datasets. The PR AUC is the area under the precision-recall curve, which plots precision against recall. The PR AUC is given by:
    
    \begin{equation}
    \text{PR AUC} = \int_{0}^{1} \text{Precision}(\text{Recall}) \, d\text{Recall .}
    \end{equation}

     The ROC AUC and PR AUC metrics are used to assess a model's performance across different thresholds. ROC AUC evaluates the trade-off between the TPR and the FPR, which is useful for understanding the model's discriminative ability across all thresholds. A higher ROC AUC indicates better performance, with a score close to 1 being ideal. In contrast, PR AUC focuses on the trade-off between precision and recall, which is particularly important in imbalanced datasets. It assesses the model's ability to correctly identify positive instances even when there are many negatives. A high PR AUC indicates that the model maintains high precision without compromising recall. Together, ROC AUC and PR AUC provide a comprehensive view of a model's effectiveness, complementing metrics like balanced  accuracy and F1 score, and helping to identify models that can make accurate predictions while meeting specific application needs.

    \section{Results: Collusion Prediction in Single Markets \label{Results I}}
    
\begin{figure}[!h]
    \centering
    \includegraphics[scale=0.4]{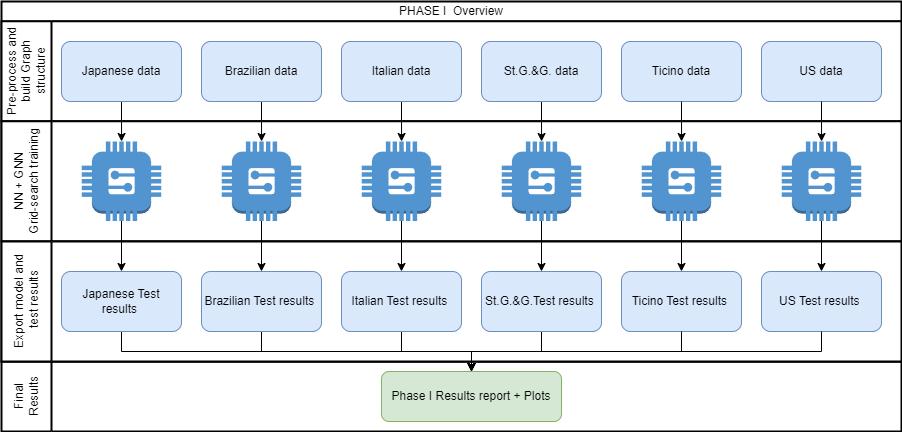}
    \caption{Phase I Overview}
    \label{fig:phaseI}
\end{figure}

        In this section, we present the performance results for training and testing the feedforward NNs and GNNs within individual country datasets. Figure \ref{fig:phaseI} presents an overview of Phase I, highlighting how the training and testing for each country takes place within that country's dataset. 

As outlined in Figure \ref{fig:method-overview}, the objective of Phase I is to train the NN and the GNN models for each country. The results from Phase I should elucidate how each network structure behaves on each type of dataset, highlighting strengths and limitations in practical scenarios.

\subsection{Phase I Results}

Table \ref{tab:trainingresults} presents the results of the two NN and GNN models trained on each dataset based on the 10 runs. Six datasets were analyzed, each corresponding to a different country: Japan, Brazil, Italy, St. Gallen and Graubünden (Switzerland), Ticino (Switzerland), and the US. For each country, the models were trained and validated on a subset of that country's data, with the remaining subset used for performance evaluation, the results of which are presented in Table \ref{tab:trainingresults}. The table shows the average metrics across the 10 runs, with the standard deviation (SD) in brackets. Bold values indicate the higher average evaluation metric value when comparing NN and GNN for each country dataset.

\begin{table}[h!]
    \caption{Test results for training within countries}
    \label{tab:trainingresults}
    \resizebox{\textwidth}{!}{%
    \begin{tabular}{ccccccc}
    \hline
    Trained on & \multicolumn{2}{c}{Japan}                   & \multicolumn{2}{c}{Brazil}                  & \multicolumn{2}{c}{Italy}                   \\ \hline
    Mean (SD)  & NN                   & GNN                  & NN                   & GNN                  & NN                   & GNN                  \\ \hline
    F1 score   & 0.40 (0.04)          & \textbf{0.70 (0.06)} & 0.71 (0.05)          & \textbf{0.81 (0.08)} & 0.55 (0.04)          & \textbf{0.57 (0.11)} \\
    Balanced accuracy   & 0.80 (0.03)          & \textbf{0.91 (0.03)} & 0.87 (0.03)          & \textbf{0.92 (0.04)} & 0.60 (0.03)          & \textbf{0.64 (0.07)} \\
    Precision  & 0.27 (0.03)          & \textbf{0.59 (0.09)} & 0.59 (0.08)          & \textbf{0.72 (0.11)} & 0.51 (0.07)          & \textbf{0.56 (0.09)} \\
    Recall     & 0.79 (0.07)          & \textbf{0.88 (0.07)} & 0.90 (0.05)          & \textbf{0.93 (0.06)} & \textbf{0.60 (0.08)} & 0.58 (0.14)          \\ \hline
    \multicolumn{1}{l}{} & \multicolumn{1}{l}{} & \multicolumn{1}{l}{} & \multicolumn{1}{l}{} & \multicolumn{1}{l}{} & \multicolumn{1}{l}{} & \multicolumn{1}{l}{} \\ \hline
    Trained on & \multicolumn{2}{c}{*St.G\&G}            & \multicolumn{2}{c}{Ticino}                  & \multicolumn{2}{c}{US}                      \\ \hline
    Mean (SD)  & NN                   & GNN                  & NN                   & GNN                  & NN                   & GNN                  \\ \hline
    F1-score   & 0.54 (0.28) & \textbf{0.62 (0.04)}          & 0.91 (0.05)          & \textbf{0.99 (0.01)} & 0.21 (0.17)          & \textbf{0.39 (0.18)} \\
    Balanced accuracy   & 0.53 (0.05)          & \textbf{0.58 (0.04)} & 0.79 (0.05)          & \textbf{0.98 (0.01)} & 0.52 (0.05)          & \textbf{0.61 (0.10)} \\
    Precision  & \textbf{0.71 (0.15)}          & 0.67 (0.05) & 0.93 (0.02)          & \textbf{1.00 (0.00)} & \textbf{0.58 (0.32)}          & 0.57 (0.18) \\
    Recall     & \textbf{0.62 (0.37)} & 0.59 (0.08)          & 0.89 (0.09)          & \textbf{0.97 (0.01)} & 0.21 (0.19)          & \textbf{0.37 (0.24)} \\ \hline
    \end{tabular}%
    }
    \footnotesize *: St. Gallen and Graubünden.
\end{table}

In Phase I, the GNN considerably outperforms the NN in terms of balanced accuracy and F1 score across all six datasets tested. When comparing the performance metrics of the NNs and GNNs across the various datasets, notable differences are observed. For the Japanese dataset, the GNN model's F1 score is 0.70, which is 30 percentage points higher than the NN's F1 score of 0.40. Similarly, the GNN's balanced accuracy is 0.91, which is 11 percentage points higher than the NN's 0.80. Precision also shows a substantial difference, with the GNN achieving 0.59 compared to 0.27 for the NN, a difference of 32 percentage points. On average, 59\% of the collusive bids predicted by the GNN model are correctly identified as collusive, compared to 27\% for the NN. Recall also shows a difference, albeit a modest one with the GNN correctly identifying 88\% of  true collusive bids compared to 79\% for the NN. 

\begin{figure}[h!]
    \centering
    
    \begin{subfigure}[b]{0.45\textwidth}
        \includegraphics[width=\textwidth]{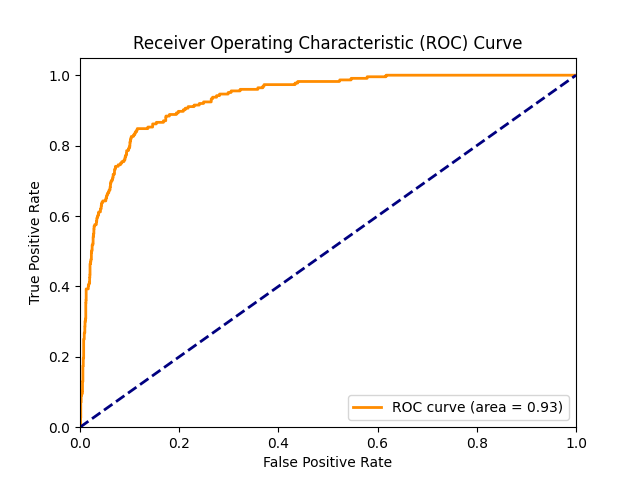}
        \caption{ROC AUC NN}
        \label{fig:a}
    \end{subfigure}
    \begin{subfigure}[b]{0.45\textwidth}
        \includegraphics[width=\textwidth]{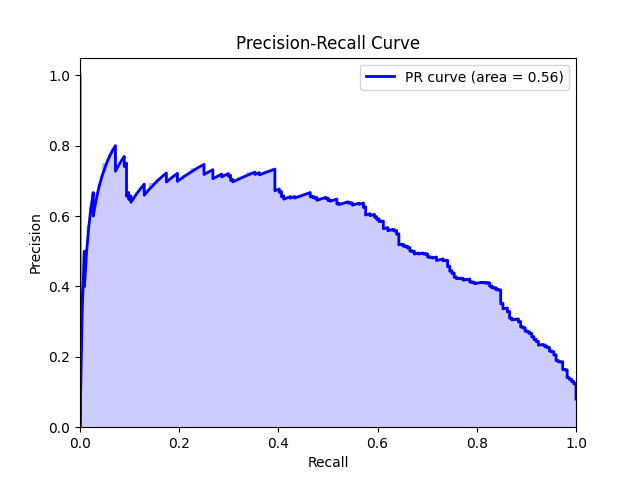 }
        \caption{PR AUC NN}
        \label{fig:b}
    \end{subfigure}
    
    \begin{subfigure}[b]{0.45\textwidth}
        \includegraphics[width=\textwidth]{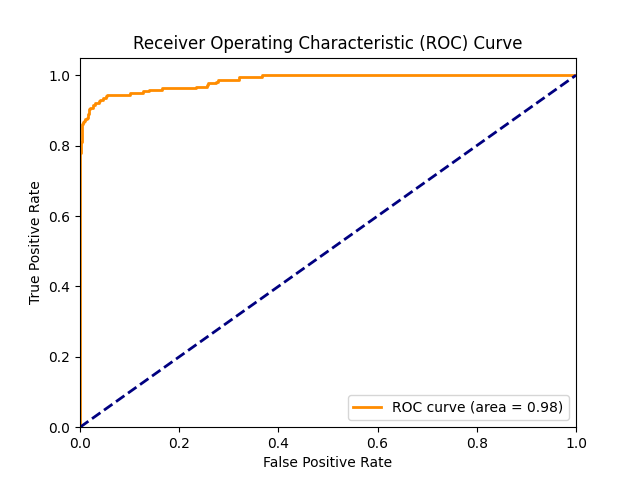}
        \caption{ROC AUC GNN}
        \label{fig:c}
    \end{subfigure}
    \begin{subfigure}[b]{0.45\textwidth}
        \includegraphics[width=\textwidth]{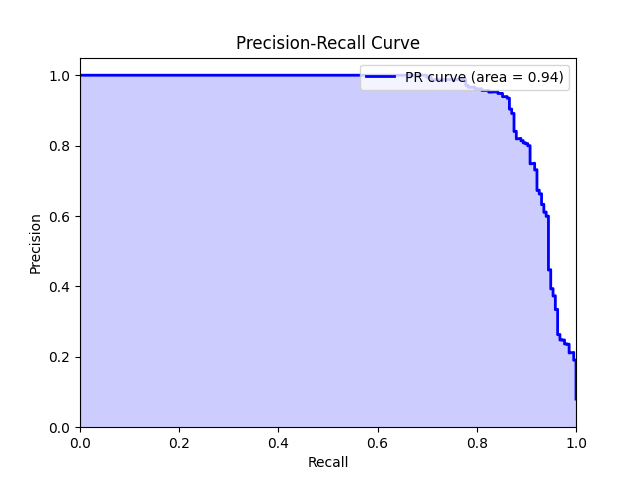}
        \caption{PR AUC GNN}
        \label{fig:d}
    \end{subfigure}
    
    \caption{ROC and PR AUC for the Japanese model - Testing set}
    \label{fig:rocprjpI}
\end{figure}

With regard to the ROC AUC and PR AUC (Figure \ref{fig:rocprjpI}), the GNN shows a moderately higher ROC AUC, at 0.98 compared to 0.93 for the NN, but a considerably higher PR AUC, at 0.94 compared to 0.56. The training process for the run with the highest F1 score can be seen in Figure \ref{fig:train-jp}, with the corresponding ROC AUC and PR AUC shown in Figure \ref{fig:rocprjpI}.

For the Brazilian dataset, the GNN model achieves an F1 score of 0.81, which is 10 percentage points higher than the NN's score of 0.71. The GNN also shows higher balanced accuracy at 0.92, compared to 0.87 with the NN, a difference of five percentage points. Precision with the GNN is 0.72, 13 percentage points higher than the NN's precision of 0.59. However, recall is only slightly higher with the GNN, at 0.93 compared to 0.90 with the NN, a difference of three percentage points. The training process for the best-performing run can be seen in Figure \ref{fig:train-br}, with the corresponding ROC AUC and PR AUC shown in Figure \ref{fig:rocprbr}.

For Italy, the GNN model achieves an F1 score of 0.57, which is two percentage points higher than the NN's score of 0.55. The GNN also shows higher balanced accuracy, at 0.64 compared to 0.60 with the NN, a difference of four percentage points. Precision with the GNN is 0.56, five percentage points higher than the NN's precision of 0.51. However, recall is slightly lower with the GNN, at 0.58 compared to 0.60 with the NN, a difference of two percentage points. The training process for the best-performing run can be seen in Figure \ref{fig:train-it}, with the corresponding ROC AUC and PR AUC shown in Figure \ref{fig:rocprit}.

In the St. Gallen and Graubünden dataset, the GNN model achieves a higher F1 score of 0.62 compared to 0.54 for the NN. Balanced accuracy is also higher for the GNN, at 0.58 compared to 0.53 for the NN, a difference of five percentage points. However, precision and recall are lower for the GNN, decreasing by four percentage points from 0.71 to 0.67 and by three percentage points from 0.62 to 0.59, respectively. The standard deviation for the F1 score, precision, and recall for the NN is relatively high at 0.28, 0.15, and 0.37, indicating greater variability across the 10 runs when compared, for example, to the models trained and tested on the Japanese, Brazilian, Italian, and Ticino datasets. The training process for the best-performing run in St. Gallen and Graubünden is shown in Figure \ref{fig:train-sw}, along with the corresponding ROC AUC and PR AUC in Figure \ref{fig:rocprsw}.

In the Ticino dataset, the GNN's F1 score is 0.99, which is eight percentage points higher than the NN's score of 0.91. Balanced accuracy is also higher with the GNN, at 0.98 compared to 0.79 with the NN, a difference of 19 percentage points. Precision with the GNN, at 1.00, is seven percentage points higher than with the NN at 0.93. Recall is also higher with the GNN, at 0.97 compared to 0.89 with the NN, a difference of eight percentage points. The training process for the best-performing run can be seen in Figure \ref{fig:train-ti}, with the corresponding ROC AUC and PR AUC curves shown in Figure \ref{fig:rocprti}.

Lastly, for the US dataset, the GNN model achieves an F1 score of 0.39, which is 18 percentage points higher than the NN's score of 0.21. Balanced accuracy is also higher with the GNN, at 0.61 compared to 0.52 with the NN, a difference of nine percentage points. While precision is slightly lower with the GNN, at 0.57 compared to 0.58 with the NN, recall is considerably higher with the GNN, which achieves 0.37 compared to 0.21 with the NN, a difference of 16 percentage points. Overall, the two models trained and tested on the US data show the lowest performance compared to those tested on the five other datasets. Additionally, the standard deviation of the evaluation metrics for both the GNN and NN models is relatively high, indicating more variability in the results over the 10 runs compared to the models for the other datasets, and the results do not appear to be stable. The training process for the best-performing run is shown in Figure \ref{fig:train-us}, with the corresponding ROC AUC and PR AUC curves shown in Figure \ref{fig:rocprus}.

These absolute improvements in the performance metrics of the GNN compared to the NN underscore that the GNNs are better at accurately predicting specific classes and identifying a higher number of relevant instances across various datasets. This demonstrates their superiority over NNs in handling the complexities of the datasets.
However, the performance of the models varies depending on the country dataset used for training. One possible reason for the low performance of both models trained on the US data could be the characteristics of the dairy market. In the US dataset, only a few companies (on average two, as shown in Table \ref{tab:collusion_data_descriptive}) participate in each tender. 
With regard to the mediocre performance of the two models trained on the Italian dataset, there are at least two possible explanations. The contracts in this dataset were awarded using the ABA method, which differs from the methods used in the other countries. In addition, seven cartels were active during the period covered by the Italian data, with a relatively large number of bids (73 on average) submitted per tender. As a result, the collusion pattern appears to be too complex to learn. The NN and GNN models trained and tested on the Ticino dataset show the highest average values for the evaluation metrics compared to the five other datasets. The Ticino dataset has the highest proportion of collusive bids (81.77\%, see Table \ref{tab:collusion_data_descriptive}) but fewer bids overall compared to the Japanese, the St. Gallen and Graubünden and the US datasets. The simpler collusion pattern in Ticino appears to be easier to learn, which might explain the strong performance of both models, although there is a possibility that the models are overfitted.

\cite{rodriguez2022collusion} trained and tested a multi-layer perceptron (MLP) on the six country datasets. It has four hidden layers with 240, 120, 70 and 35 neurons, and they performed 500 iterations on four model settings. Among other metrics, they report the average balanced accuracy for the 500 runs in each setting. The MLP on the Brazilian dataset achieved a balanced accuracy of 71.3\% to 72.4\%, while the Italian, Japanese, Ticino, St. Gallen and Graubünden, and US datasets all had balanced accuracies around 50\%. Compared to these results, the NN in our study performs better, which may be due to differences in network architecture, hyperparameters or input features.

    \section{Results: Transportability \label{Results II}}
    
\begin{figure}[!h]
    \centering
    \includegraphics[scale=0.4]{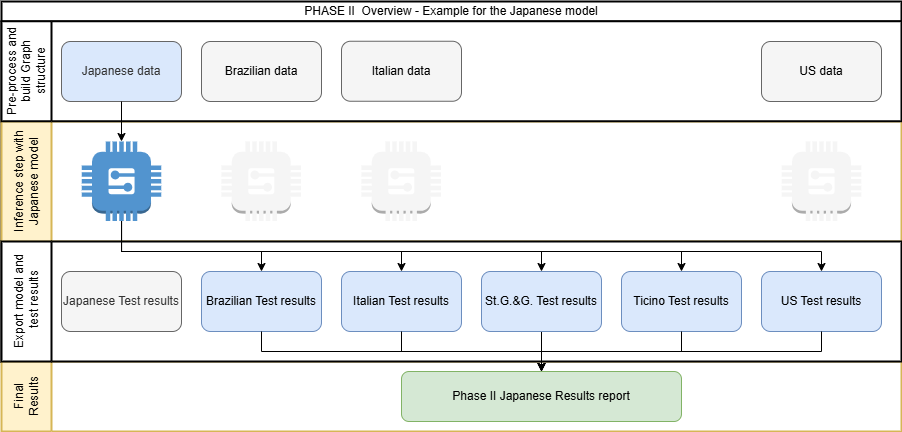}
    \caption{Phase II Overview}
    \label{fig:phaseII}
\end{figure}

        In Part II of our study, we analyze how well the models trained in Phase I generalize when applied to data from other countries without further fine-tuning. In other words, we test each model as a zero-shot learner. Figure \ref{fig:phaseII} presents a simple overview of the Japanese case in Phase II, showing how the model trained in Phase I is applied to the other country datasets. The same process is used for all other datasets except for the two from Switzerland because they have only one edge type available for training, limiting their flexibility for evaluation on other datasets.

\subsection{Phase II Results}

In contrast to Phase I, in which we split the data from each country into training, validation and test sets, Phase II involves taking the two models (i.e., GNN and NN) trained in Phase I for each country (except Switzerland) and applying them separately to the complete datasets of the other countries (including the two Swiss regions), which thus serve as test datasets to evaluate the models. This is done for all 10 runs. These test datasets exclude the data used to train a model because including this data would bias model inference. Hence, we present 10 model comparisons for each of the four countries considered. 

The results for the NN and GNN models trained on the Japanese data are presented in Table \ref{tab:inf-jp}. The average evaluation metrics from 10 runs are shown, with the standard deviation (SD) in brackets. Values in bold indicate the higher average evaluation metric when comparing the performance of the Japanese NN model to that of the Japanese GNN model when applied to the test dataset from a different country. 

\begin{table}[h!]
    \caption{Inference with model trained on Japanese data}
    \label{tab:inf-jp}
    \resizebox{\textwidth}{!}{%
    \begin{tabular}{ccccccc}
    \hline
    \multicolumn{7}{c}{{\color[HTML]{333333} \textbf{Trained on Japanese data}}}                                                                                     \\ \hline
    Inference on & \multicolumn{2}{c}{Japan}                   & \multicolumn{2}{c}{Brazil}                             & \multicolumn{2}{c}{Italy}                   \\ \hline
    Mean (SD)  & NN           & GNN                  & NN                   & GNN                                       & NN                   & GNN                  \\ \hline
    F1 score   & -          & -                     & \textbf{0.01 (0.01)}  & 0.00 (0.00)                               & 0.06 (0.02)          & \textbf{0.24 (0.19)} \\
    Balanced accuracy   & -  & -                    & 0.50 (0.00)           & 0.50 (0.00)                       & \textbf{0.50 (0.00)}          & 0.48 (0.01) \\
    Precision  & -          & -                     & $1.00^\dagger$ (0.00)          & $1.00^\dagger$ (0.00)    & \textbf{0.47 (0.02)}          & 0.43 (0.20) \\
    Recall     & -          & -                     & $0.00^\dagger$ (0.01)          & $0.00^\dagger$ (0.00)   & 0.03 (0.01)            & \textbf{0.28 (0.27)}          \\ \hline
    \multicolumn{1}{l}{} & \multicolumn{1}{l}{} & \multicolumn{1}{l}{} & \multicolumn{1}{l}{} & \multicolumn{1}{l}{} & \multicolumn{1}{l}{} & \multicolumn{1}{l}{} \\ \hline
    Inference on & \multicolumn{2}{c}{*St.G\&G}                                 & \multicolumn{2}{c}{Ticino}                        & \multicolumn{2}{c}{US}                      \\ \hline
    Mean (SD)  & NN                             & GNN                       & NN                   & GNN                            & NN                          & GNN                  \\ \hline
    F1 score   & 0.00 (0.00)                    & \textbf{0.56 (0.10)}       & 0.05 (0.02)          & \textbf{0.76 (0.09)}          & 0.00 (0.00)                & \textbf{0.15 (0.10)} \\
    Balanced accuracy   & 0.50 (0.00)           & \textbf{0.56 (0.02)}       & 0.51 (0.00)          & \textbf{0.78 (0.03)}          & 0.50 (0.00)               & \textbf{0.53 (0.03)} \\
    Precision  & $\textbf{1.00}^\dagger$ \textbf{(0.00)}  & 0.67 (0.05)       & \textbf{0.99 (0.01)}          & 0.98 (0.03) & \textbf{0.90 (0.30)}          & 0.27 (0.13) \\
    Recall     & $0.00^\dagger$ (0.00)              & \textbf{0.53 (0.21)}       & 0.03 (0.01)          & \textbf{0.64 (0.15)}      & 0.00 (0.00)             & \textbf{0.26 (0.32)} \\ \hline
    \end{tabular}%
    }
    \footnotesize \({\dagger}\): Precision = 1 together with recall = 0 indicates that the model classifies all observations as non-collusive.
        
    \footnotesize *: St. Gallen and Graubünden.
\end{table}

Concerning F1 scores, the Japanese GNN model outperforms the Japanese NN in four out of five test datasets, with particularly notable differences observed when either of the Swiss datasets is used as the test dataset. The Japanese NN model's F1 score is close to 0 in all tests, reaching a maximum of 6\% when the Italian dataset is used as the test dataset. In terms of precision and recall, some models, indicated by \({\dagger}\), predict all instances as non-collusive, as evidenced by precision scores of 1 and recall scores of 0. The Japanese GNN correctly identifies some portion of collusive and non-collusive cases in four out of five test datasets, compared to three out of five for the Japanese NN. For the US test dataset, the Japanese GNN achieves an F1 score of 0.15 compared to 0.00 for the NN. Although this performance is very low, the Japanese GNN model provides a more balanced prediction across classes, with non-zero predictions for collusive and non-collusive bids. 

Interestingly, both Japanese models classify almost all Brazilian bids as non-collusive, indicating very low predictive power. The prediction performance of both Japanese models is low for the Italian and the US test datasets but the Japanese GNN performs better than the Japanese NN. Moreover, the results of the Japanese GNN when using Italian test dataset are slightly better than its results with the US test dataset. Lastly, the Japanese GNN model performs substantially better than the Japanese NN model when either of the Swiss datasets (St. Gallen or Graubünden and Ticino) is used as a test dataset, with the GNN model achieving an F1 score of 0.05 compared to 0.00 with the NN, and 0.76 compared to 0.56 with the NN, respectively. The Swiss and Italian data involve tenders in the construction sector, similar to the Japanese data on which the two models were trained. The better performance of the Japanese NN and GNN on these country test datasets suggests a correlation with the data source and type.

When considering the results for the NN model and GNN model trained with the Brazilian data, we observe that, in terms of the F1 score, the Brazilian NN outperforms the Brazilian GNN in three out of five cases (Table \ref{tab:inf-br}). Unlike the Japanese models, which see better results for the Japanese GNN than for the Japanese NN when using either of the Swiss test datasets, the Brazilian models see better results for the Brazilian NN when the St. Gallen and Graubünden data are used as the test dataset, with the model achieving an F1 score of 0.72 compared 0.36 for the Brazilian GNN. However, the results for balanced accuracy are almost the same for the two Brazilian models when tested on the Swiss datasets, with a value of 0.51 for the Brazilian NN and 0.49 for the Brazilian GNN. Overall, both Brazilian models show rather low predictive performance regardless of the test dataset used. The collusive patterns in tenders for oil infrastructure projects in Brazil would seem to differ too much from those in the other country datasets.

\begin{table}[h!]
    \caption{Inference with model trained on Brazilian data}
    \label{tab:inf-br}
    \resizebox{\textwidth}{!}{%
    \begin{tabular}{ccccccc}
    \hline
    \multicolumn{7}{c}{{\color[HTML]{333333} \textbf{Trained on Brazilian data}}}                                                                                     \\ \hline
    Inference on & \multicolumn{2}{c}{Japan}                        & \multicolumn{2}{c}{Brazil}           & \multicolumn{2}{c}{Italy}                   \\ \hline
    Mean (SD)  & NN           & GNN                                 & NN            & GNN                  & NN                   & GNN                  \\ \hline
    F1 score   & \textbf{0.14 (0.04)}          & 0.12 (0.05)        & -             & -                    & \textbf{0.11 (0.02)}          & 0.06 (0.08) \\
    Balanced accuracy   & \textbf{0.51 (0.08)}  & 0.47 (0.06)       & -             & -                     & \textbf{0.51 (0.00)}          & 0.50 (0.01) \\
    Precision  & \textbf{0.08 (0.02)}          & 0.07 (0.03)        & -             & -                     & \textbf{0.45 (0.00)}          & 0.33 (0.11) \\
    Recall     & 0.52 (0.23)          & \textbf{0.63 (0.35)}        & -             & -                     & \textbf{0.06 (0.01)}            &0.04 (0.05)          \\ \hline
    \multicolumn{1}{l}{} & \multicolumn{1}{l}{} & \multicolumn{1}{l}{} & \multicolumn{1}{l}{} & \multicolumn{1}{l}{} & \multicolumn{1}{l}{} & \multicolumn{1}{l}{} \\ \hline
    Inference on & \multicolumn{2}{c}{*St.G\&G}                                 & \multicolumn{2}{c}{Ticino}                        & \multicolumn{2}{c}{US}                      \\ \hline
    Mean (SD)  & NN                             & GNN                       & NN                   & GNN                            & NN                          & GNN                  \\ \hline
    F1 score   & \textbf{0.72 (0.03)}           & 0.36 (0.28)       & 0.00 (0.00)          & \textbf{0.13 (0.28)}          & 0.11 (0.03)                & \textbf{0.21 (0.03)} \\
    Balanced accuracy   & \textbf{0.51 (0.01)}  & 0.49 (0.02)       & 0.49 (0.00)          & \textbf{0.51 (0.05)}          & 0.44 (0.01)               & \textbf{0.47 (0.03)} \\
    Precision  & \textbf{0.60 (0.01)}              & 0.53 (0.07)       & 0.07 (0.15)          & \textbf{0.61 (0.30)}    & 0.09 (0.01)          & \textbf{0.13 (0.01)} \\
    Recall     & \textbf{0.90 (0.08)}              & 0.39 (0.36)       & 0.00 (0.00)          & \textbf{0.12 (0.29)}      & 0.16 (0.06)             & \textbf{0.64 (0.25)} \\ \hline
    \end{tabular}%
    }
    \footnotesize \({\dagger}\): Precision = 1 together with recall = 0 indicates that the model classifies all observations as non-collusive.
        
    \footnotesize *: St. Gallen and Graubünden.
\end{table}

When analyzing the models trained on the Italian data, we observe that the Italian NN outperforms the Italian GNN in terms of F1 score in four out of five test datasets (Japan, Brazil, St. Gallen and Graubünden, and Ticino), as shown in Table \ref{tab:inf-it}. However, the differences in F1 score between the two models are not substantial. Overall, the predictive performance of both Italian models is low for each of the five test datasets. This is not surprising given their performance in Phase I, in which both models also showed low predictive power when tested with the Italian test set.

\begin{table}[h!]
    \caption{Inference with model trained on Italian data}
    \label{tab:inf-it}
    \resizebox{\textwidth}{!}{%
    \begin{tabular}{ccccccc}
    \hline
    \multicolumn{7}{c}{{\color[HTML]{333333} \textbf{Trained on Italian data}}}                                                                                     \\ \hline
    Inference on & \multicolumn{2}{c}{Japan}                        & \multicolumn{2}{c}{Brazil}                    & \multicolumn{2}{c}{Italy}                   \\ \hline
    Mean (SD)  & NN           & GNN                                 & NN                        & GNN                  & NN         & GNN                  \\ \hline
    F1 score   & \textbf{0.14 (0.02)}          & 0.10 (0.06)        & \textbf{0.42 (0.03)}      & 0.30 (0.08)          & -         & - \\
    Balanced accuracy   & 0.50 (0.05)           & 0.50 (0.03)       & \textbf{0.67 (0.03)}      & 0.51 (0.07)           & -         & - \\
    Precision  & \textbf{0.08 (0.01)}          & 0.07 (0.03)        & \textbf{0.27 (0.02)}      & 0.18 (0.05)             & -          & - \\
    Recall     & \textbf{0.60 (0.17)}          & 0.43 (0.40)        & \textbf{0.92 (0.07)}      & 0.81 (0.25)            & -            & -         \\ \hline
    \multicolumn{1}{l}{} & \multicolumn{1}{l}{} & \multicolumn{1}{l}{} & \multicolumn{1}{l}{} & \multicolumn{1}{l}{} & \multicolumn{1}{l}{} & \multicolumn{1}{l}{} \\ \hline
    Inference on & \multicolumn{2}{c}{*St.G\&G}                                 & \multicolumn{2}{c}{Ticino}                        & \multicolumn{2}{c}{US}                      \\ \hline
    Mean (SD)  & NN                             & GNN                       & NN                   & GNN                            & NN                          & GNN                  \\ \hline
    F1 score   & \textbf{0.59 (0.06)}           & 0.41 (0.31)            & \textbf{0.52 (0.17)}          & 0.48 (0.34)          & 0.22 (0.02)                & \textbf{0.23 (0.05)} \\
    Balanced accuracy   & 0.48 (0.01)           & \textbf{0.49 (0.03)}  & \textbf{0.57 (0.09)}          & 0.55 (0.09)                & 0.50 (0.01)               & 0.50 (0.03) \\
    Precision  & 0.57 (0.01)              & \textbf{0.58 (0.15)}       & 0.85 (0.06)          & \textbf{0.87 (0.06)}             & 0.14 (0.01)          & 0.14 (0.02) \\
    Recall     & \textbf{0.60 (0.10)}              & 0.48 (0.40)       & 0.40 (0.17)          & \textbf{0.44 (0.37)}      & 0.49 (0.08)             & \textbf{0.74 (0.29)} \\ \hline
    \end{tabular}%
    }
    \footnotesize \({\dagger}\): Precision = 1 together with recall = 0 indicates that the model classifies all observations as non-collusive.
        
    \footnotesize *: St. Gallen and Graubünden.
\end{table}

\begin{table}[h!]
    \caption{Inference with model trained on US data}
    \label{tab:inf-us}
    \resizebox{\textwidth}{!}{%
    \begin{tabular}{ccccccc}
    \hline
    \multicolumn{7}{c}{{\color[HTML]{333333} \textbf{Trained on US data}}}                                                                                     \\ \hline
    Inference on & \multicolumn{2}{c}{Japan}                        & \multicolumn{2}{c}{Brazil}                    & \multicolumn{2}{c}{Italy}                   \\ \hline
    Mean (SD)  & NN           & GNN                                 & NN                        & GNN                  & NN         & GNN                  \\ \hline
    F1 score   & 0.07 (0.06)          & \textbf{0.13 (0.04)}        & 0.06 (0.08)      & \textbf{0.24 (0.09)}          & 0.30 (0.22)         & \textbf{0.54 (0.05)} \\
    Balanced accuracy   & 0.48 (0.03)  & \textbf{0.50 (0.02)}       & 0.38 (0.09)      & \textbf{0.44 (0.05)}           & 0.49 (0.02)         & \textbf{0.50 (0.02)} \\
    Precision  & \textbf{0.34 (0.43)}          & 0.08 (0.01)        & \textbf{0.34 (0.44)}      & 0.15 (0.05)             & \textbf{0.56 (0.29)}          & 0.39 (0.02) \\
    Recall     & 0.26 (0.29)          & \textbf{0.79 (0.32)}        & 0.13 (0.22)      & \textbf{0.70 (0.28)}            & 0.39 (0.33)            & \textbf{0.89 (0.16)}         \\ \hline
    \multicolumn{1}{l}{} & \multicolumn{1}{l}{} & \multicolumn{1}{l}{} & \multicolumn{1}{l}{} & \multicolumn{1}{l}{} & \multicolumn{1}{l}{} & \multicolumn{1}{l}{} \\ \hline
    Inference on & \multicolumn{2}{c}{*St.G\&G}                                 & \multicolumn{2}{c}{Ticino}                        & \multicolumn{2}{c}{US}                      \\ \hline
    Mean (SD)  & NN                             & GNN                       & NN                   & GNN                            & NN                          & GNN                  \\ \hline
    F1 score   & 0.27 (0.24)           & \textbf{0.43 (0.24)}               & 0.39 (0.30)          & \textbf{0.53 (0.33)}          & -                & - \\
    Balanced accuracy   & 0.52 (0.02)           & \textbf{0.54 (0.02)}       & 0.45 (0.05)          & 0.45 (0.15)                & -               & - \\
    Precision  & \textbf{0.76 (0.16)}              & 0.61 (0.16)            & \textbf{0.82 (0.14)}          & 0.71 (0.27)             & -          & - \\
    Recall     & 0.23 (0.25)              & \textbf{0.40 (0.26)}            & 0.33 (0.28)          & \textbf{0.50 (0.36)}      & -             & - \\ \hline
    \end{tabular}%
    }
    \footnotesize \({\dagger}\): Precision = 1 together with recall = 0 indicates that the model classifies all observations as non-collusive.
        
    \footnotesize *: St. Gallen and Graubünden.
\end{table}

The GNN model trained on the US data outperforms the NN model trained on the US data in terms of F1 score in all five test datasets. However, the US GNN model tested with the Italian test dataset achieves the highest average F1 score of 0.54, indicating low performance (Table \ref{tab:inf-us}). The values for balanced accuracy are almost the same for the two US models when tested on the five test datasets. This overall result is not surprising because the models trained and tested on the US data in Phase I showed the lowest predictive performance.

In summary, the Japanese GNN model tested on the Ticino test dataset shows the best performance, with an average F1 score of 0.76, balanced accuracy of 0.78, precision of 0.98 and recall of 0.64 (Table \ref{tab:inf-jp}). In addition, the standard deviation over the 10 runs is low at 0.09, 0.03, 0.03 and 0.15. The results therefore do not vary substantially over the 10 runs and appear to be stable. Both countries' data contains tenders in the construction sector. The collusion patterns learned by the GNN in the Japanese data appear to be similar to those in the Ticino data.

Overall, the predictive power for both the NN and GNN models in Phase II is rather low across all test datasets. It appears that the individual markets are too different in terms of their collusive patterns.

    \section{Final Remarks \label{finalremarks}}
    
        This study evaluated the performance of NNs and GNNs for collusion detection in different national markets and for different tender types. Our investigation shows the  superiority of GNNs over their NN counterparts in metrics such as F1 score, balanced accuracy, precision, and recall, especially in Phase I of our study, in which we trained and tested both models on the same country dataset.
The differing performance between GNNs and NNs, particularly in terms of precision and F1 score, underscores the importance of model selection and the choice of an appropriate deep learning architecture. The interconnectivity of data points and relational patterns in data relevant to collusion detection favor GNNs.  
Moreover, our study provides insights into the scenarios in which GNNs are particularly useful, especially in relation to the density and configuration of relational linkages.
We also found that the type of tender plays a crucial role in the modeling of collusion detection. For instance, tenders for the procurement of school milk in the US dataset are less complex, with fewer participating companies. As a result, GNNs trained on the US dataset show a lower scale of connectivity, making the benefits of GNNs less pronounced. Conversely, in more complex collusive environments, such as the Italian dataset, in which tenders were awarded according to the ABA method and seven cartels operated simultaneously, the GNN could not effectively learn the patterns with the available information. Additional information about the participating companies could potentially improve the GNN's ability to learn and uncover collusive patterns.
In summary, our findings show that GNNs are beneficial for predicting collusion and can outperform standard feedforward NNs. We hope this study can serve as a starting point for further applications of GNNs in other fields of economics.  
In future research, we plan to build on our findings by exploring recent architectures, such as attention-based graph neural networks, in conjunction with even more collusion datasets, with the aim of understanding better which data attributes favor the use of GNNs.

    \section*{Disclosure statement}
    
        The authors declare no conflict of interest with the publication of this manuscript. 
        
    \newpage

    \bibliographystyle{elsarticle-harv} 
%\bibliography{article.bib}

    % Appendix
    \newpage
    \appendix
    \fancyhf{}
    \fancyhead[R]{\leftmark}
    \pagestyle{fancy}

    \section{Extract of Raw Data\label{Appendix}}
        \begin{figure}[!hp]
        \centering
        \includegraphics[scale=0.6, angle=90]{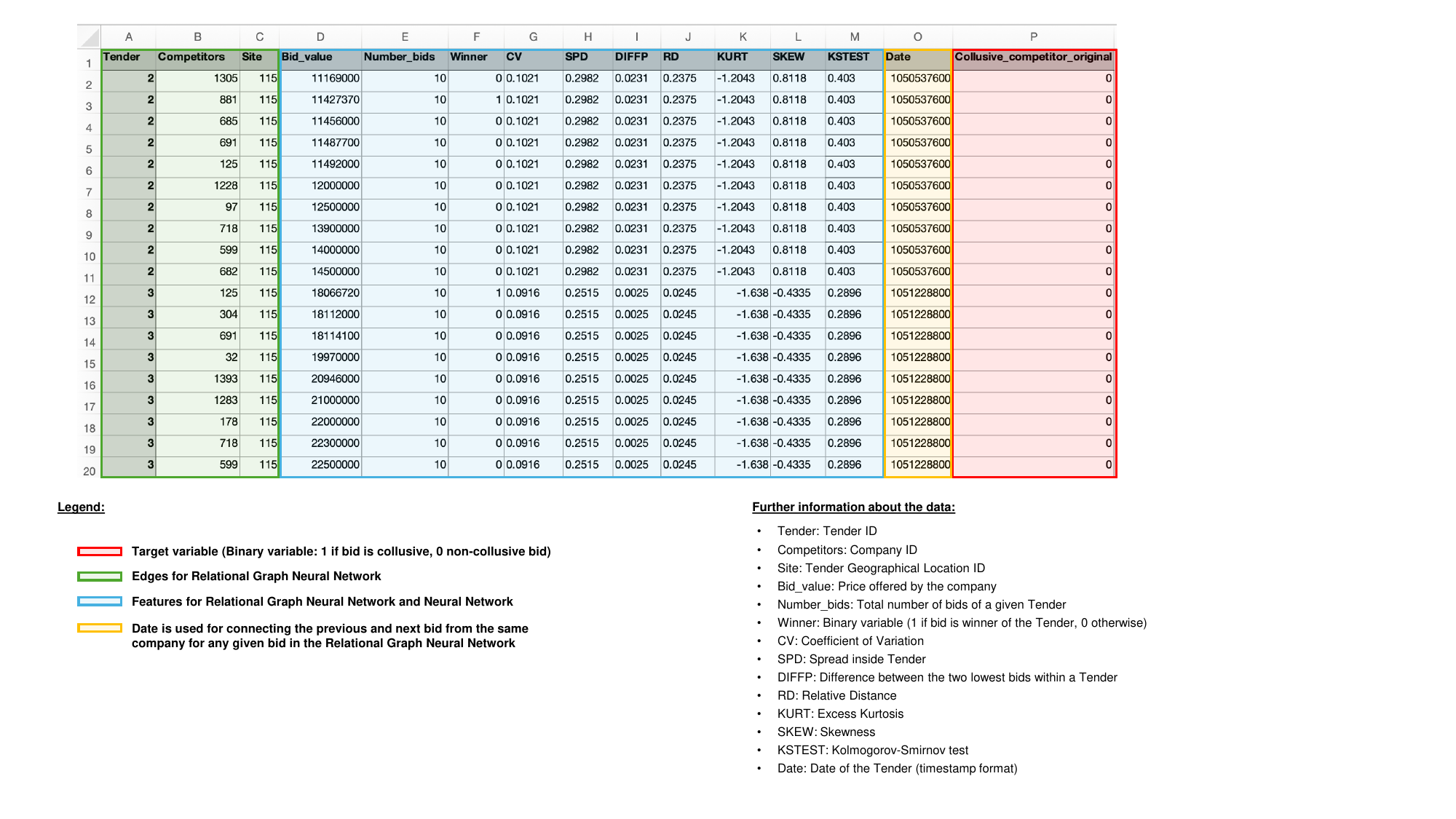}
        \caption{Extract of the raw data of Japan (tabular data).}
        \label{fig:rawdata_japan}
    \end{figure}

        \clearpage
        
    \newpage
    \section{Screening Variables\label{Appendix-b}}
    
        \begin{itemize}
    \item \textbf{Coefficient of Variation (CV)}: The coefficient of variation is a statistical measure of the dispersion of data points in a data series around the mean. It is defined as the ratio of the standard deviation of bids within a tender $s_t$ to the mean bid value within a tender $\overline{bid}_t$.
    \[
    CV_t=\frac{s_t}{\overline{bid}_t}
    \]

    \item \textbf{Spread inside Tender (SPD)}: This measure indicates the spread or variability of bids within a tender. It can be calculated as the relative difference between the highest and lowest bids.
    \[
    SPD_t=\frac{max(bid_t) - min(bid_t)}{min(bid_t)}
    \]

    \item \textbf{Difference between the two lowest bids within a Tender (DIFFP)}: This is a specific measure of competitiveness within a tender, focusing on the gap between the second lowest bid $bid_{2t}$ and lowest bid within a tender.
    \[
    DIFFP_t=\frac{bid_{2t}-min(bid_t)}{min(bid_t)}
    \]

    \item \textbf{Relative Distance (RD)}: Relative distance is a similar screen to DIFFP which uses the standard deviation of losing bids within a tender $s_{losingbids,t}$ instead of the minimum bid in the denominator.
    \[
    RD_t=\frac{bid_{2t}-min(bid_t)}{s_{losingbids,t}}
    \]

    \item \textbf{Excess Kurtosis (KURT)}: Excess kurtosis is a measure of the ``tailedness" of the probability distribution of a real-valued random variable. It is the kurtosis minus 3, which adjusts the value to make the normal distribution have an excess kurtosis of 0.
    \[
    KURT_t = \frac{n_t(n_t+1)}{(n_t-1)(n_t-2)(n_t-3)} \sum \left( \frac{bid_{it}-\overline{bid}_t}{s_t} \right)^4 - \frac{3(n_t-1)^2}{(n_t-2)(n_t-3)},
    \]
    where $n_t$ is the number of bids in a given tender. \\

    \item \textbf{Skewness (SKEW)}: Skewness is a measure of the asymmetry of the probability distribution of a real-valued random variable about its mean. Positive skew indicates a distribution with an asymmetric tail extending towards more positive values.
    \[
    SKEW_t=\frac{n_t}{(n_t-1)(n_t-2)} \sum \left( \frac{bid_{it} - \overline{bid}_t}{s_t} \right)^3
    \]

    \item \textbf{Kolmogorov-Smirnov Test (KSTEST)}: The Kolmogorov-Smirnov test is a nonparametric test used to determine if the distribution of bid values in a tender follows a uniform distribution. The test statistic \(D\) is the maximum absolute difference between the cumulative distribution function of the bid values in a tender $F_{n_t}(bid_t)$ and the uniform cumulative distribution function $F(bid_t)$.
    \[
    D_t = \sup_{bid_t} = |F_n(bid_t) - F(bid_t)|
    \]
\end{itemize}

        \clearpage

    \newpage
    \section{Distribution of Variables\label{Appendix-c}}
    
            \begin{figure}[!ht]
      \centering
      \makebox[\textwidth][c]{%
        \includegraphics[width=1.0\textwidth]{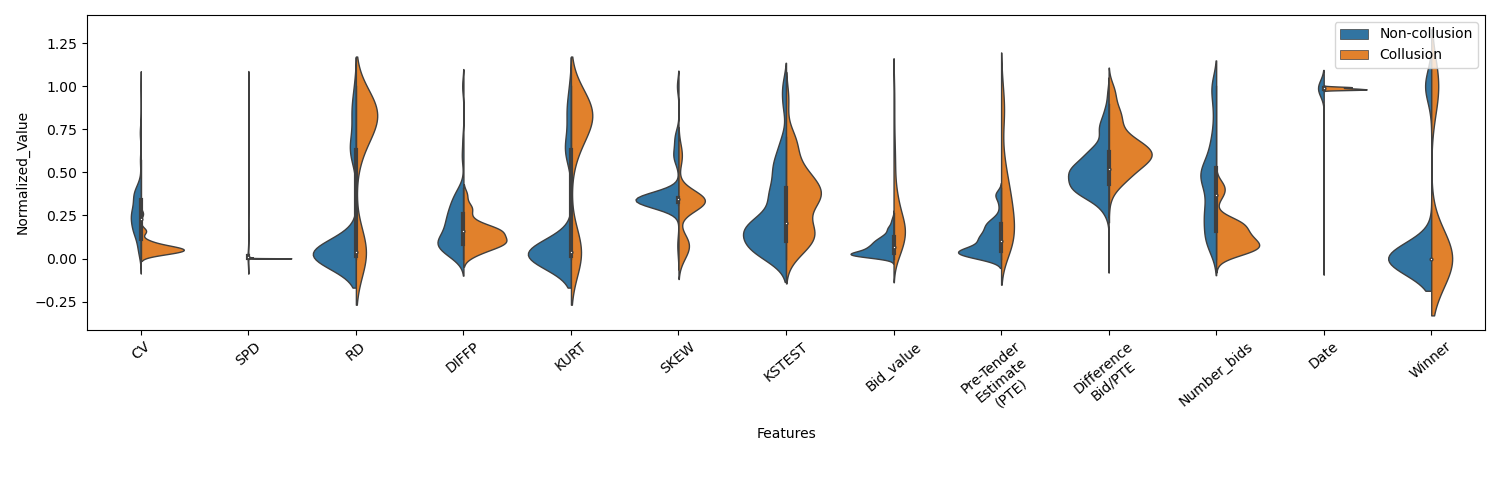}
      }
      \caption{Distribution of variables among collusive and non-collusive bids in the Brazilian dataset.}
      \label{fig:variables_comparison_br}
    \end{figure}

        \begin{figure}[!ht]
      \centering
      \makebox[\textwidth][c]{%
        \includegraphics[width=1.0\textwidth]{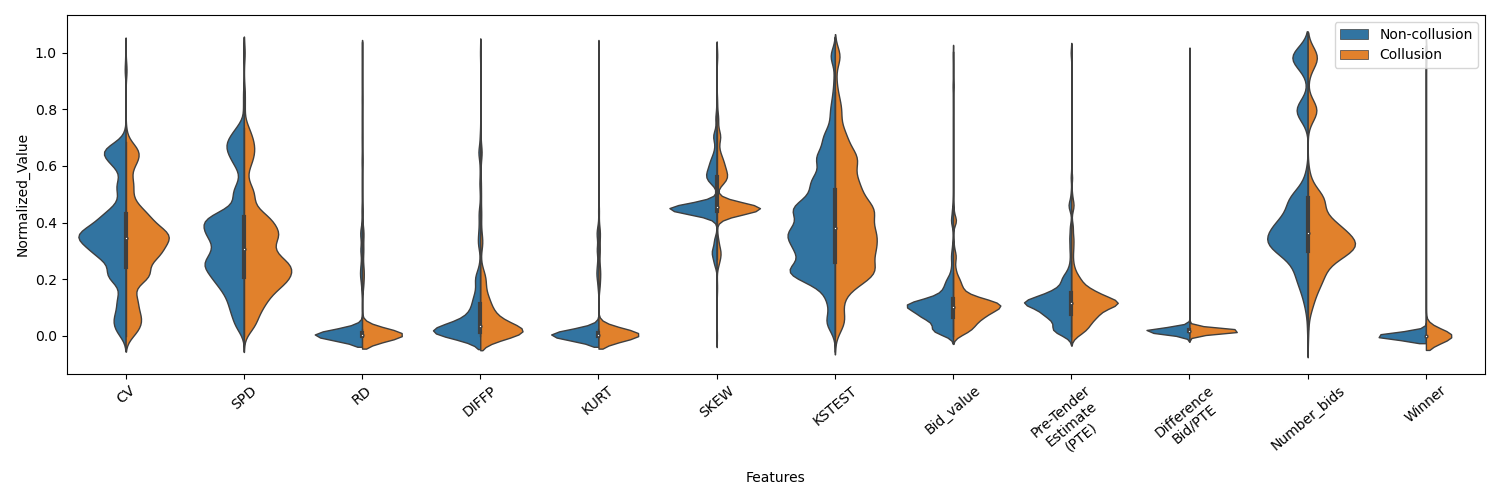}
      }
      \caption{Distribution of variables among collusive and non-collusive bids in the Italian dataset.}
      \label{fig:variables_comparison_it}
    \end{figure}

            \begin{figure}[!ht]
      \centering
      \makebox[\textwidth][c]{%
        \includegraphics[width=1.0\textwidth]{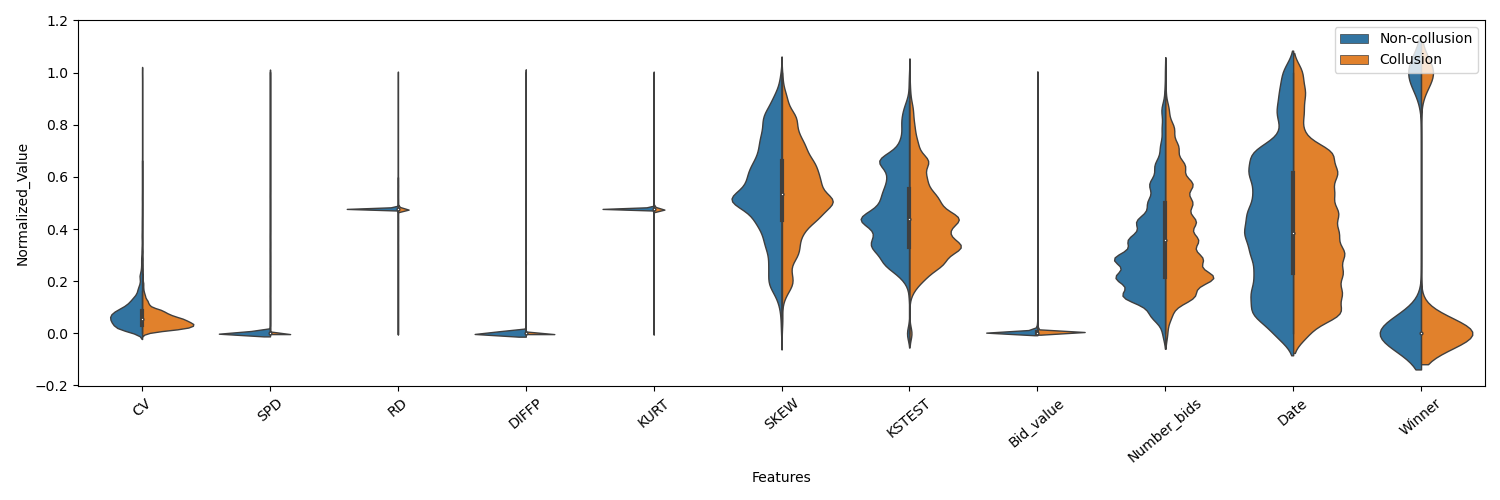}
      }
      \caption{Distribution of variable among collusive and non-collusive bids in the St. Gallen and Graubünden dataset.}
      \label{fig:variables_comparison_sw}
    \end{figure}

                \begin{figure}[!ht]
      \centering
      \makebox[\textwidth][c]{%
        \includegraphics[width=1.0\textwidth]{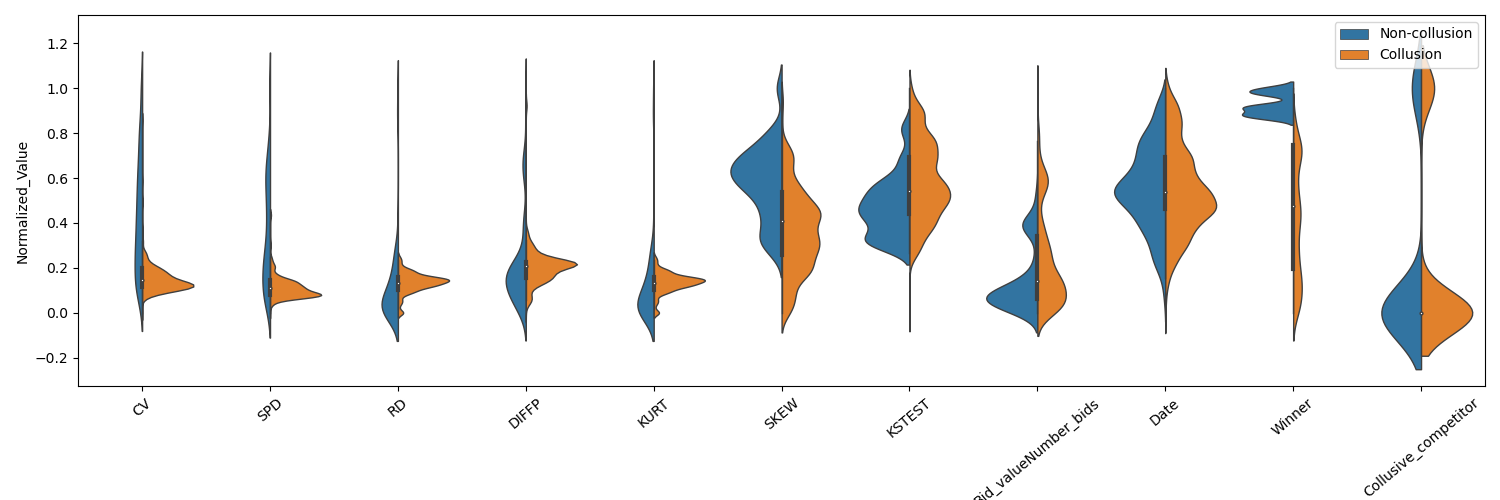}
      }
      \caption{Distribution of variables among collusive and non-collusive bids in the Ticino dataset.}
      \label{fig:variables_comparison_ti}
    \end{figure}

                    \begin{figure}[!ht]
      \centering
      \makebox[\textwidth][c]{%
        \includegraphics[width=1.0\textwidth]{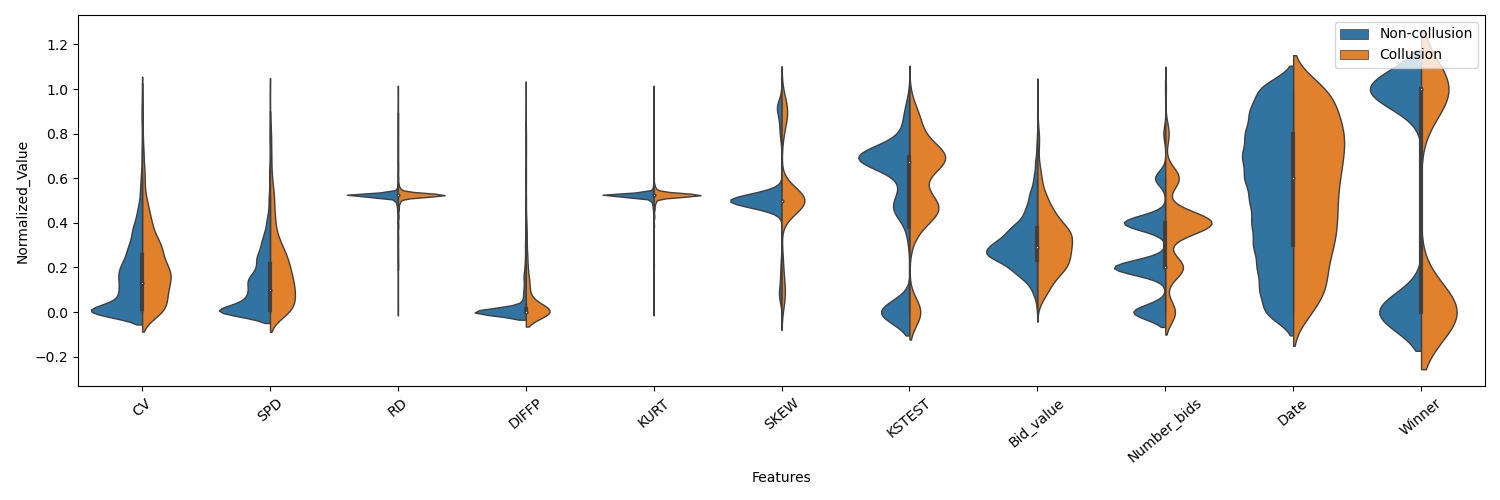}
      }
      \caption{Distribution of variables among collusive and non-collusive bids in the US dataset.}
      \label{fig:variables_comparison_us}
    \end{figure}
        \clearpage

    \newpage
    \section{Descriptive Analysis of Global Datasets\label{Appendix-d}}
    
        \begin{table}[h]
    \captionsetup{justification=centering} 
    \caption{Observations, percentage of collusive bids and average number of bids per auction}
    \label{tab:collusion_data_descriptive}
    \begin{tabular}{lccc}
    \hline
    \textbf{Dataset} & \textbf{Number of bids} & \textbf{\makecell{Percentage of \\ collusive bids}} & \textbf{\makecell{Average number \\ of bids per auction}} \\
    \hline
    Brazil & 683 & 18.74\% & 6.76 \\
    Italy & 20286 & 39.86\% & 72.97 \\
    Japan & 13515 & 8.09\% & 12.51 \\
    St. Gallen and Graubünden* & 21231 & 58.88\% & 4.89 \\
    Ticino* & 1629 & 81.77\% & 7.27 \\
    America & 7004 & 12.36\% & 1.91 \\
    \hline
    \end{tabular}
    \footnotesize *Both St. Gallen and Graubünden and Ticino are regions (cantons) in Switzerland.
\end{table}
        \clearpage
        
    \newpage
    \section{F1 Score and Loss during Training\label{Appendix-e}}
    
        \begin{figure}[h!]
    \centering
    \begin{subfigure}{\textwidth}
        \centering
        \includegraphics[width=0.9\linewidth]{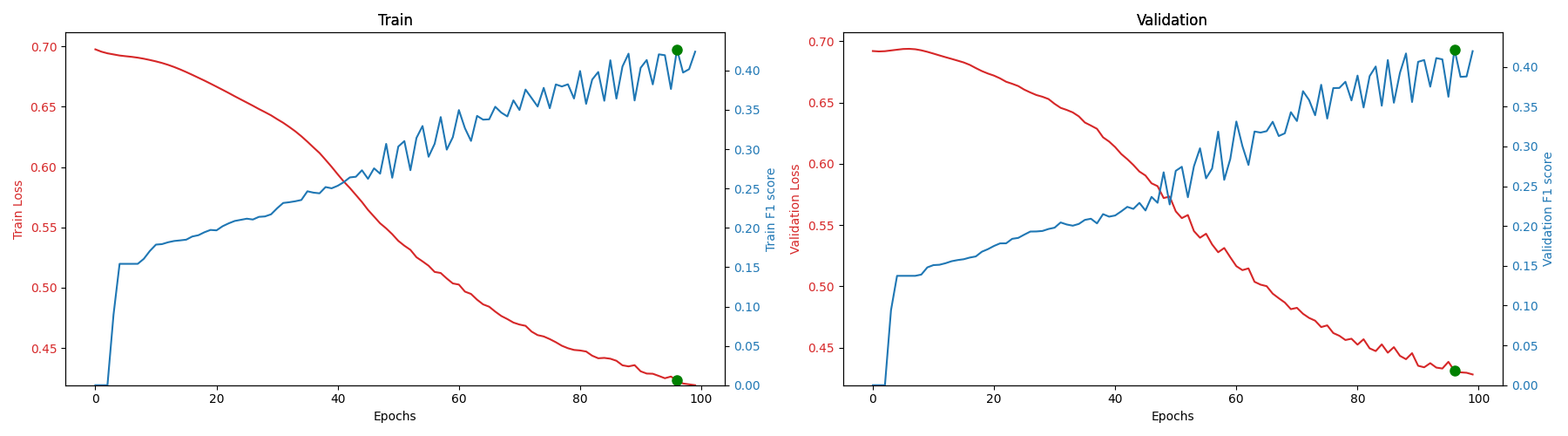}
        \caption{NN model F1 score vs Loss}
    \end{subfigure}
    \vspace{0.5cm} 
    \begin{subfigure}{\textwidth}
        \centering
        \includegraphics[width=0.9\linewidth]{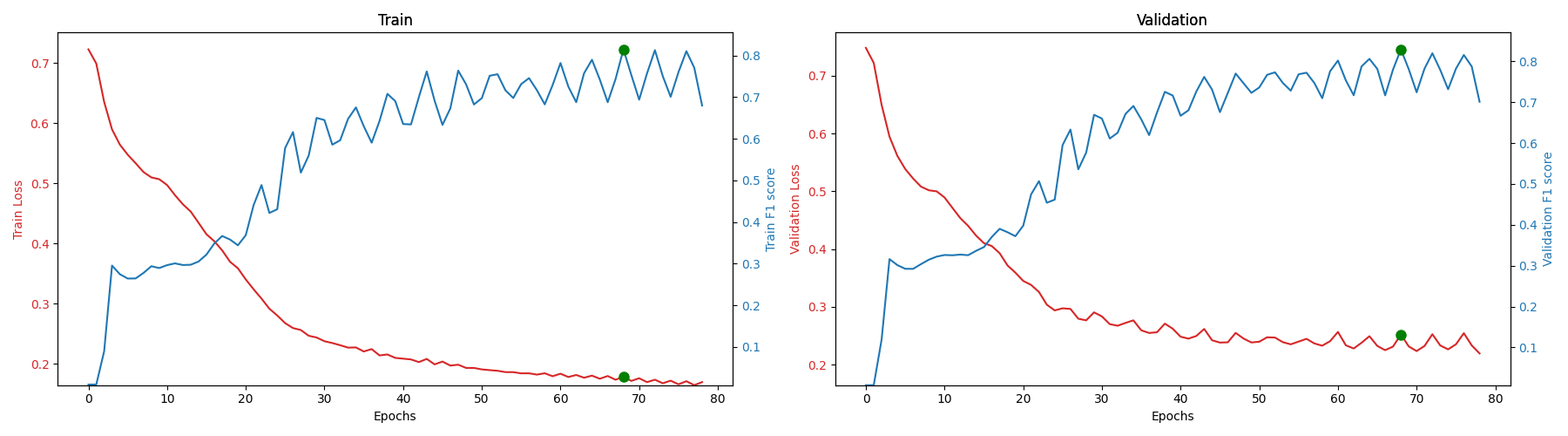} 
        \caption{GNN model F1 score vs Loss}
    \end{subfigure}
    \caption{Training progress for the Japanese model} 
    \label{fig:train-jp}
\end{figure}

\begin{figure}[h!]
    \centering
    \begin{subfigure}{\textwidth}
        \centering
        \includegraphics[width=0.9\linewidth]{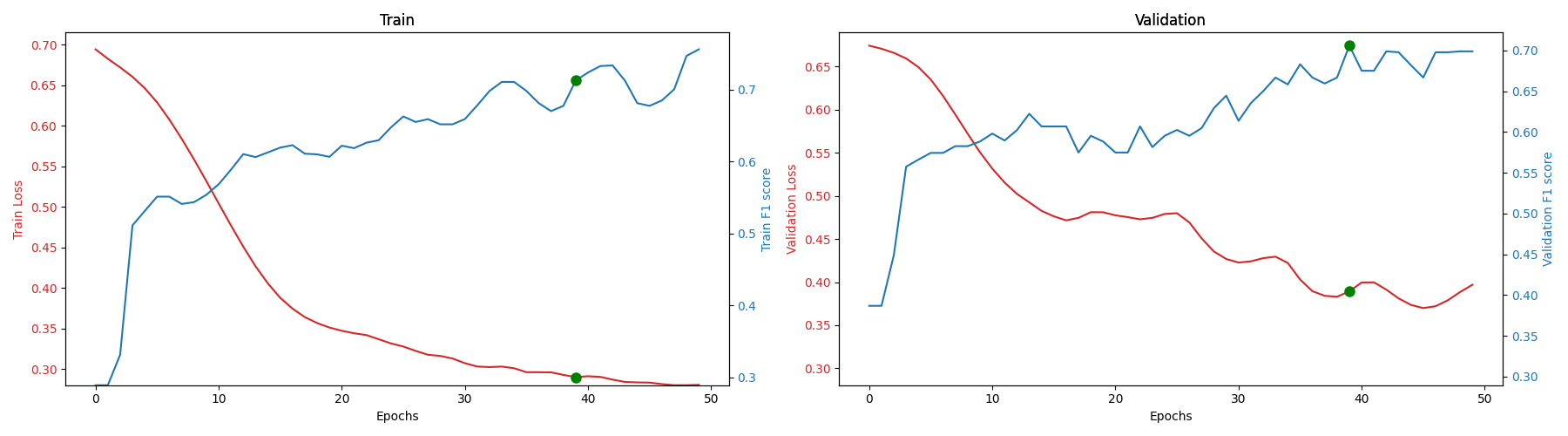}
        \caption{NN model F1 score vs Loss}
    \end{subfigure}
    \vspace{0.5cm} 
    \begin{subfigure}{\textwidth}
        \centering
        \includegraphics[width=0.9\linewidth]{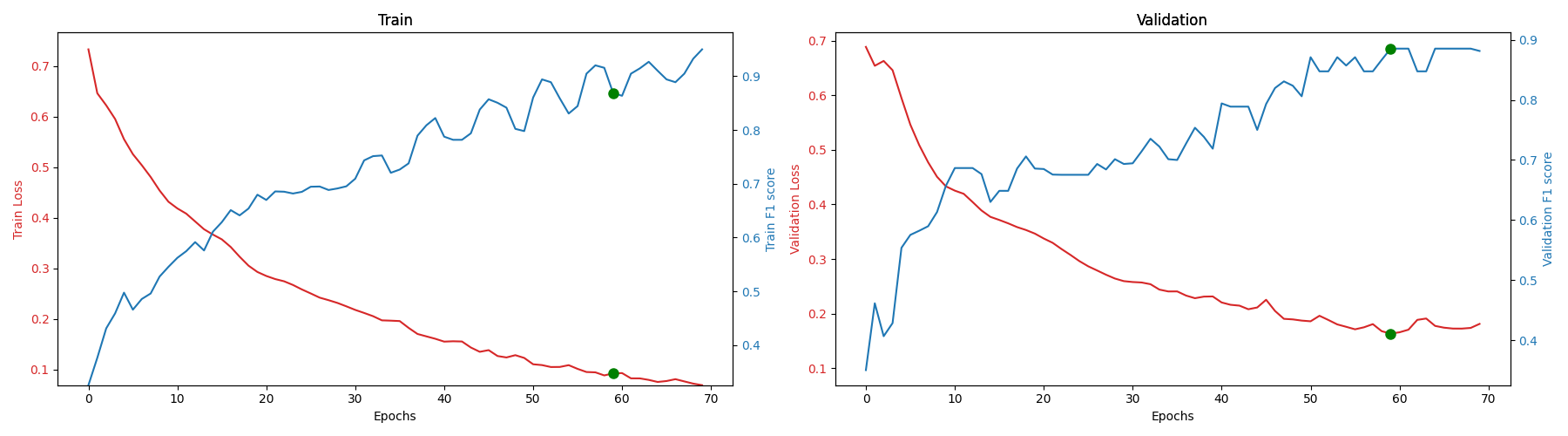} 
        \caption{GNN model F1 score vs Loss}
    \end{subfigure}
    \caption{Training progress for the Brazilian model} 
    \label{fig:train-br}
\end{figure}

\begin{figure}[h!]
    \centering
    \begin{subfigure}{\textwidth}
        \centering
        \includegraphics[width=0.9\linewidth]{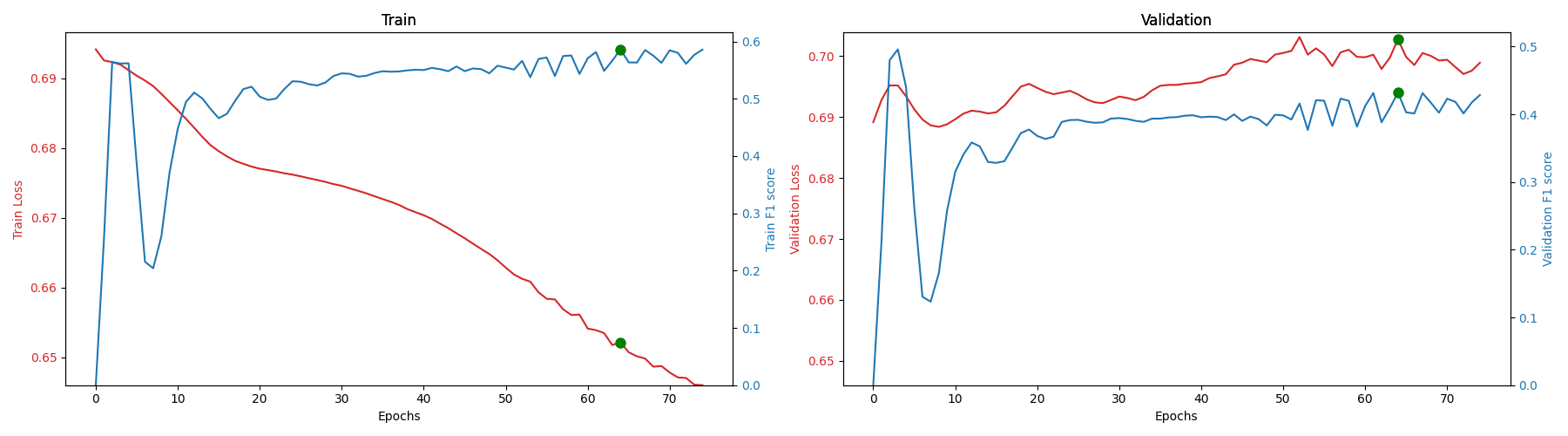}
        \caption{NN model F1 score vs Loss}
    \end{subfigure}
    \vspace{0.5cm} 
    \begin{subfigure}{\textwidth}
        \centering
        \includegraphics[width=0.9\linewidth]{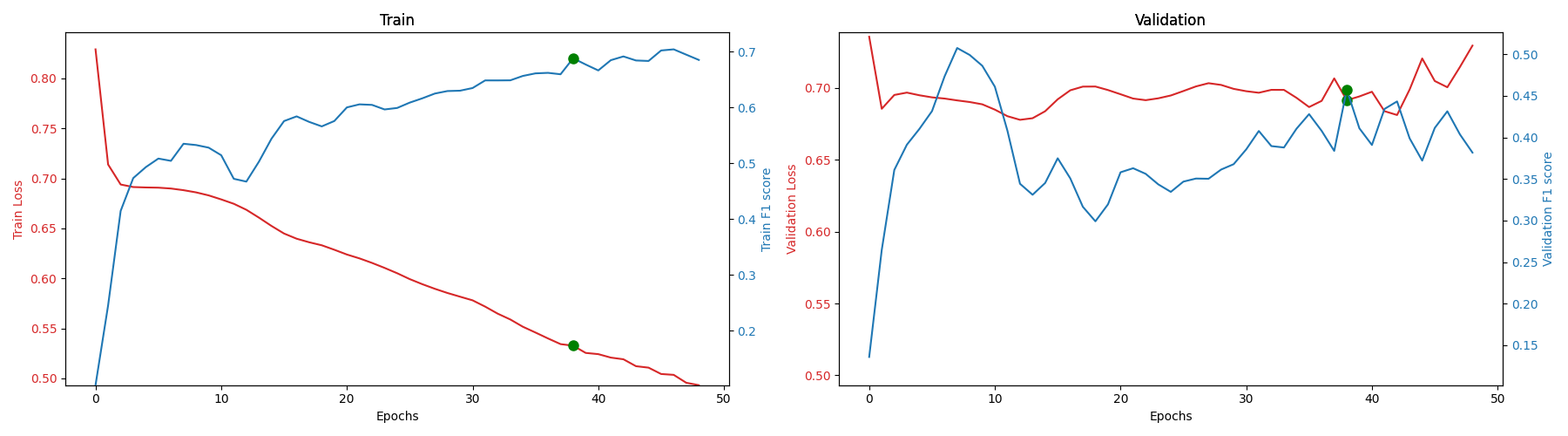} 
        \caption{GNN model F1 score vs Loss}
    \end{subfigure}
    \caption{Training progress for the Italian model} 
    \label{fig:train-it}
\end{figure}

\begin{figure}[h!]
    \centering
    \begin{subfigure}{\textwidth}
        \centering
        \includegraphics[width=0.9\linewidth]{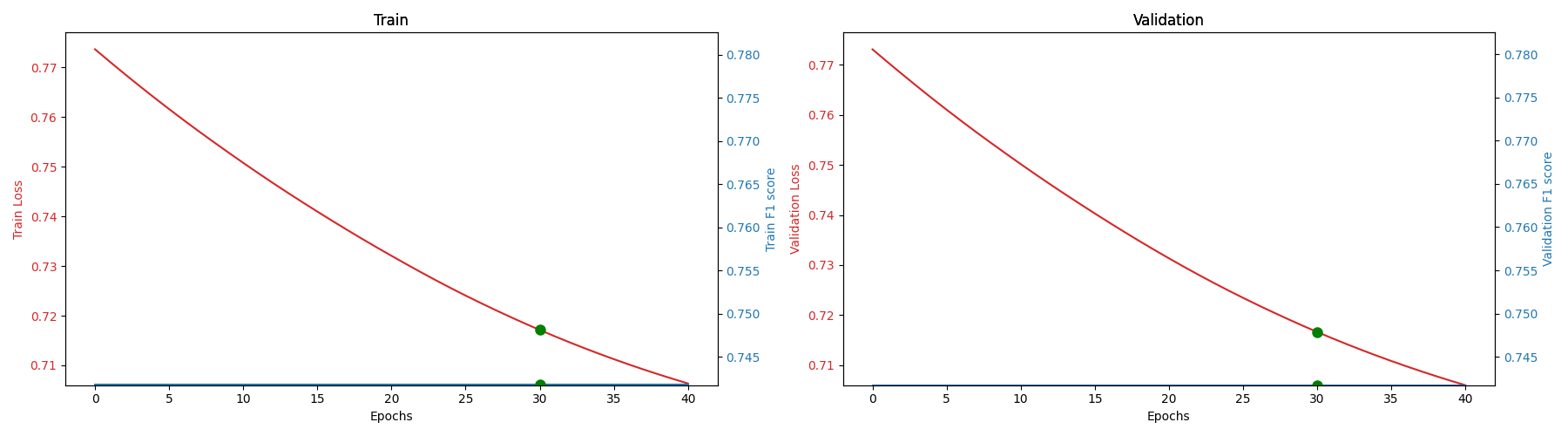}
        \caption{NN model F1 score vs Loss}
    \end{subfigure}
    \vspace{0.5cm} 
    \begin{subfigure}{\textwidth}
        \centering
        \includegraphics[width=0.9\linewidth]{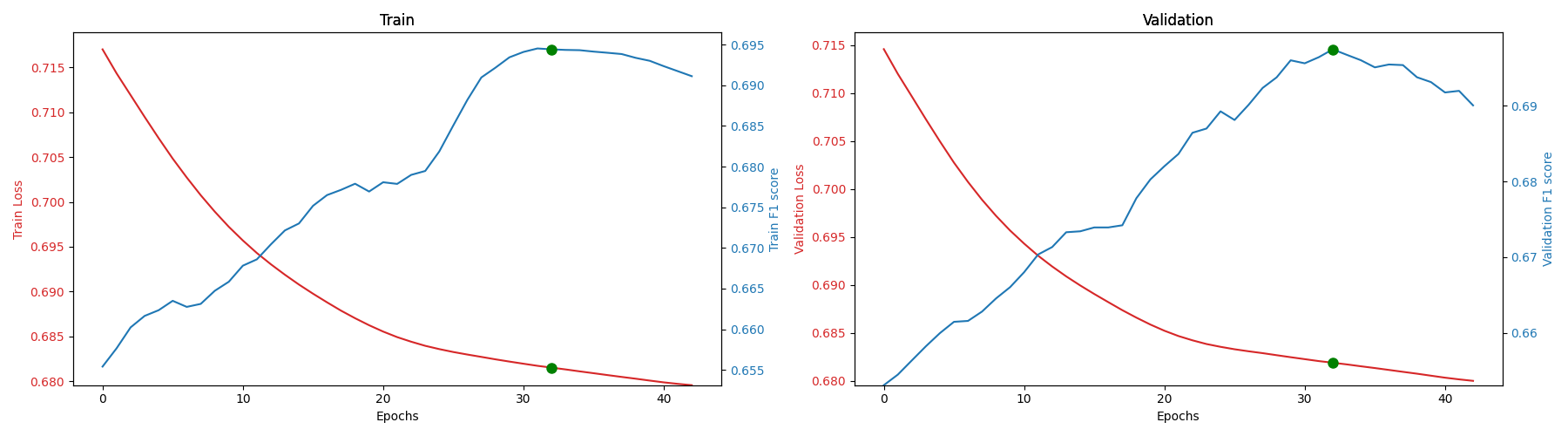} 
        \caption{GNN model F1 score vs Loss}
    \end{subfigure}
    \caption{Training progress for the St. Gallen and Graubünden model} 
    \label{fig:train-sw}
\end{figure}

\begin{figure}[h!]
    \centering
    \begin{subfigure}{\textwidth}
        \centering
        \includegraphics[width=0.9\linewidth]{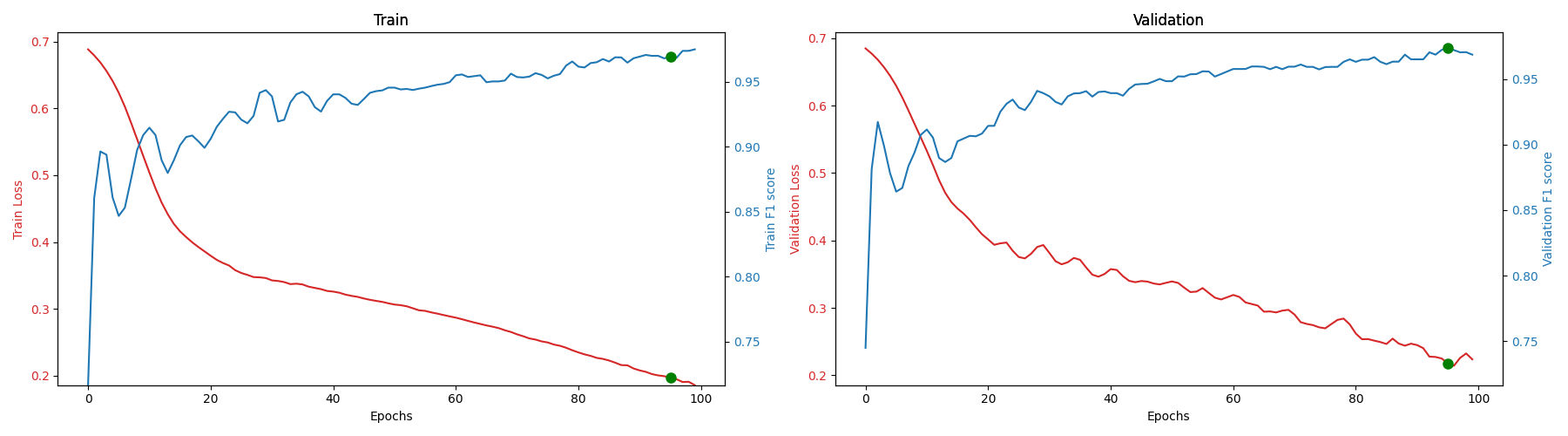}
        \caption{NN model F1 score vs Loss}
    \end{subfigure}
    \vspace{0.5cm} 
    \begin{subfigure}{\textwidth}
        \centering
        \includegraphics[width=0.9\linewidth]{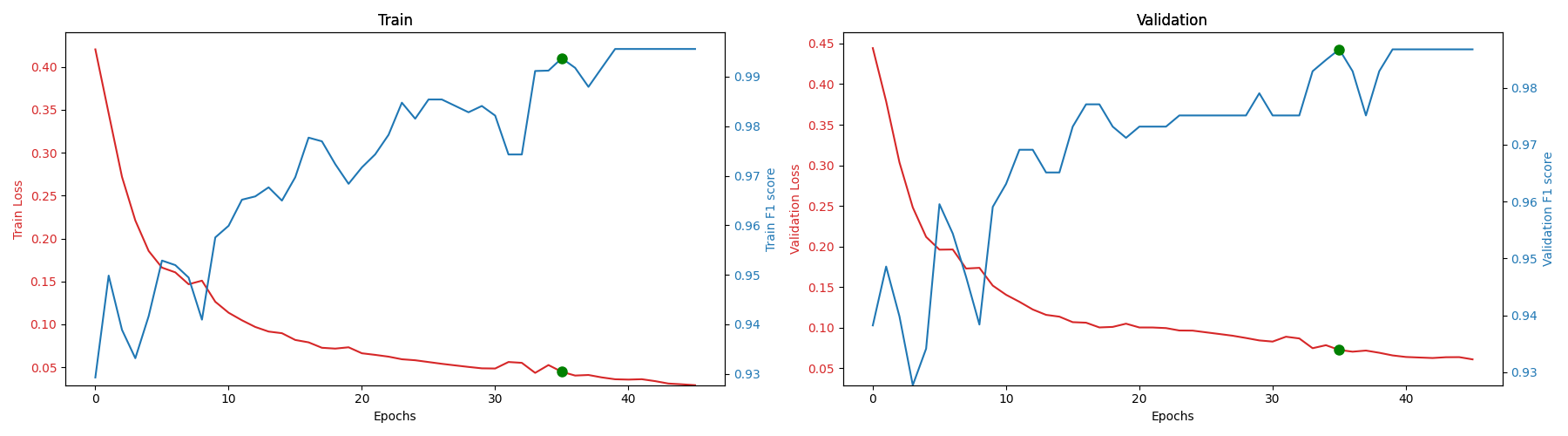} 
        \caption{GNN model F1 score vs Loss}
    \end{subfigure}
    \caption{Training progress for the Ticino model} 
    \label{fig:train-ti}
\end{figure}

\begin{figure}[h!]
    \centering
    \begin{subfigure}{\textwidth}
        \centering
        \includegraphics[width=0.9\linewidth]{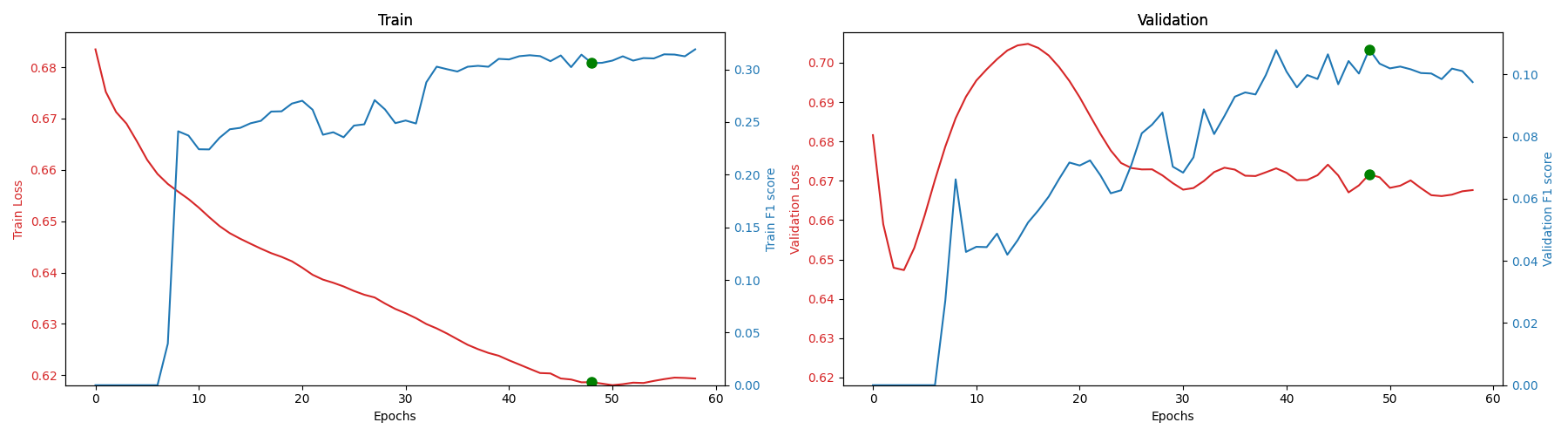}
        \caption{NN model F1 score vs Loss}
    \end{subfigure}
    \vspace{0.5cm} 
    \begin{subfigure}{\textwidth}
        \centering
        \includegraphics[width=0.9\linewidth]{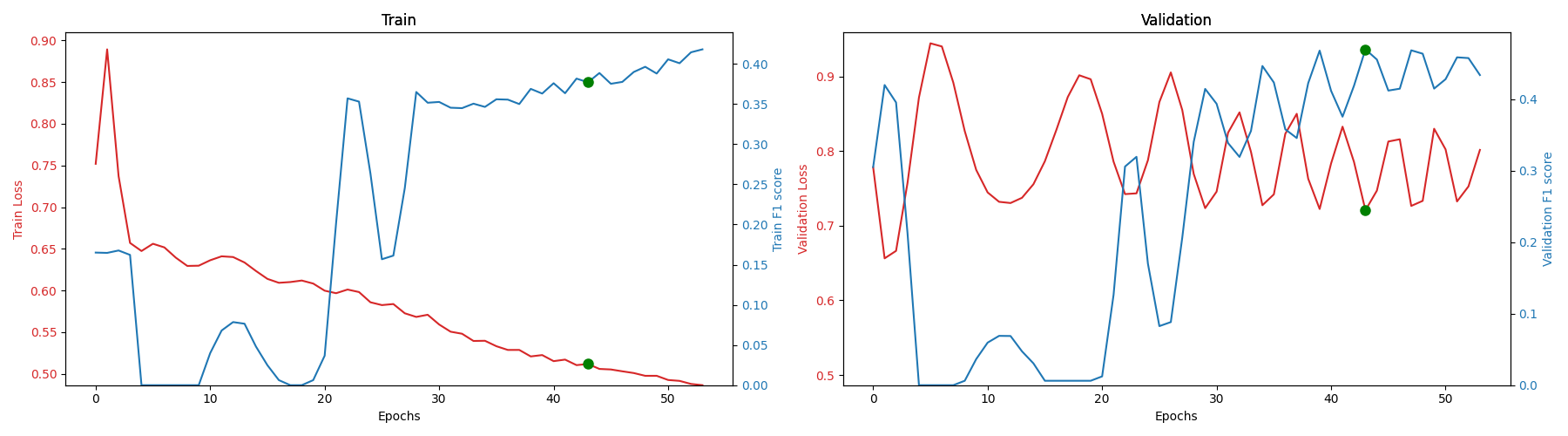} 
        \caption{GNN model F1 score vs Loss}
    \end{subfigure}
    \caption{Training progress for the US model} 
    \label{fig:train-us}
\end{figure}
        \clearpage

    \newpage
    \section{ROC and PR AUC\label{Appendix-f}}
    
        \begin{figure}[htp]
    \centering
    
    \begin{subfigure}[b]{0.45\textwidth}
        \includegraphics[width=\textwidth]{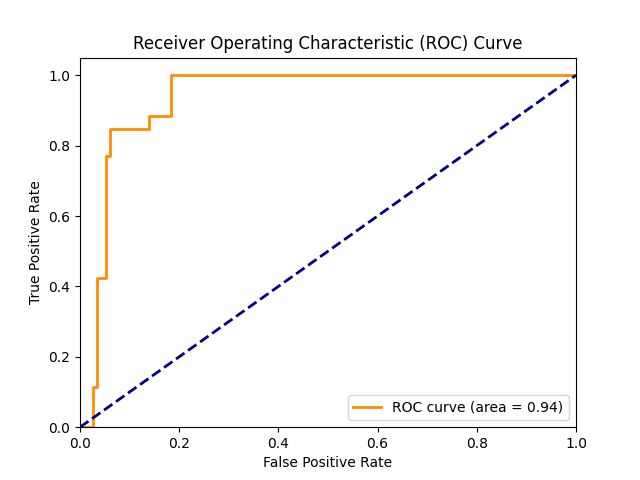}
        \caption{ROC AUC NN}
    \end{subfigure}
    \hfill 
    \begin{subfigure}[b]{0.45\textwidth}
        \includegraphics[width=\textwidth]{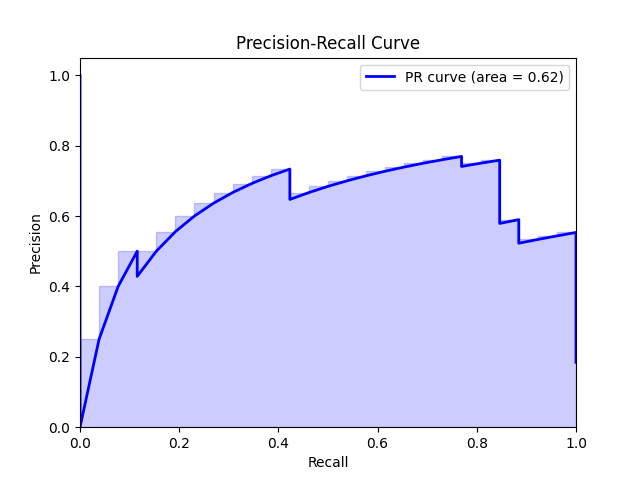 }
        \caption{PR AUC NN}
    \end{subfigure}
    
    \begin{subfigure}[b]{0.45\textwidth}
        \includegraphics[width=\textwidth]{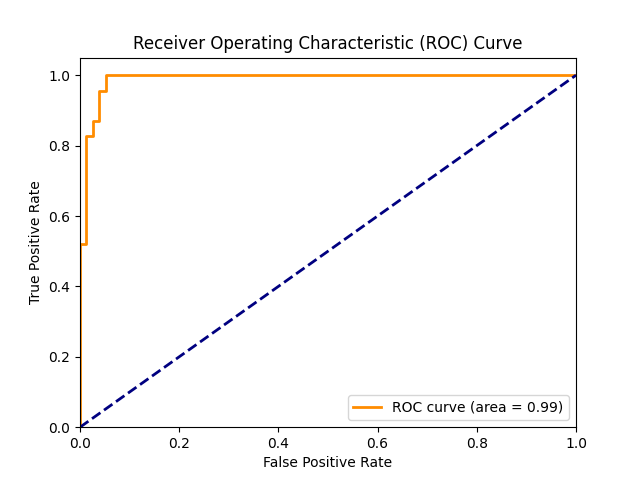}
        \caption{ROC AUC GNN}
    \end{subfigure}
    \hfill 
    \begin{subfigure}[b]{0.45\textwidth}
        \includegraphics[width=\textwidth]{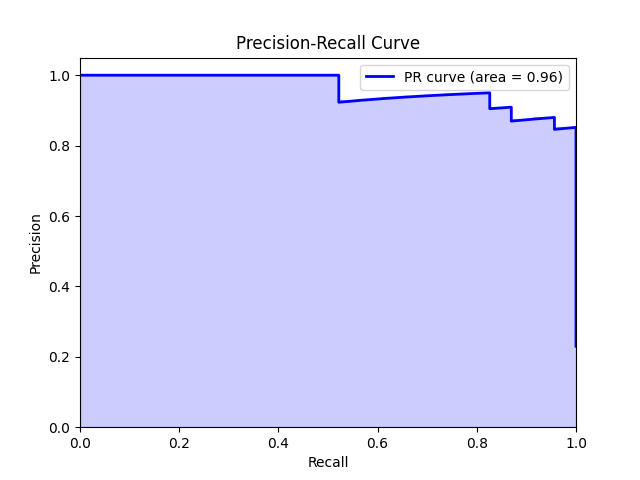}
        \caption{PR AUC GNN}
    \end{subfigure}
    
    \caption{ROC and PR AUC for the Brazilian model - Testing set}
    \label{fig:rocprbr}
\end{figure}

\begin{figure}[htp]
    \centering
    
    \begin{subfigure}[b]{0.45\textwidth}
        \includegraphics[width=\textwidth]{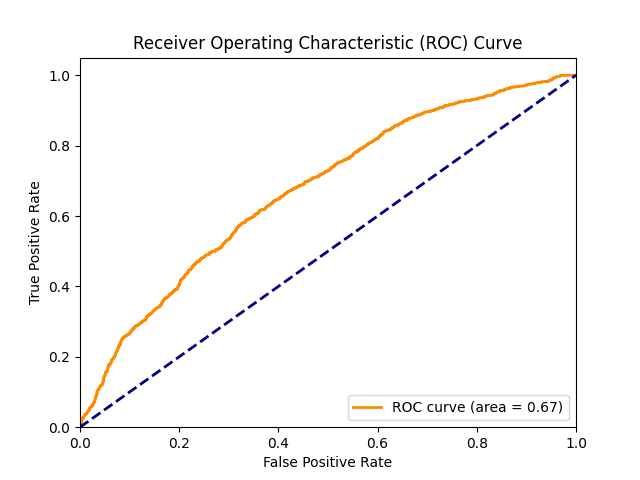}
        \caption{ROC AUC NN}
    \end{subfigure}
    \hfill 
    \begin{subfigure}[b]{0.45\textwidth}
        \includegraphics[width=\textwidth]{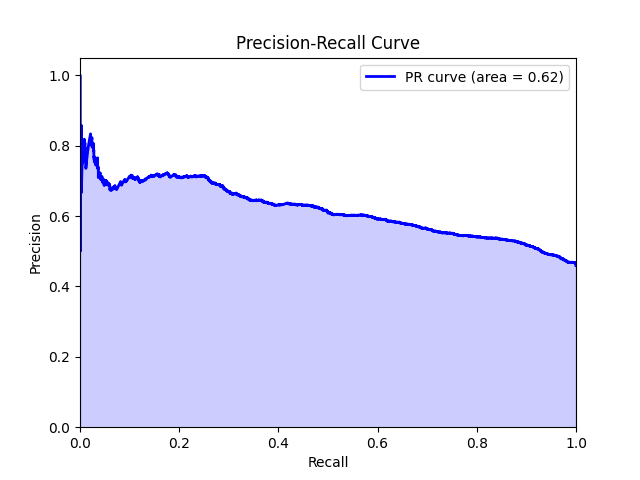 }
        \caption{PR AUC NN}
    \end{subfigure}
    
    \begin{subfigure}[b]{0.45\textwidth}
        \includegraphics[width=\textwidth]{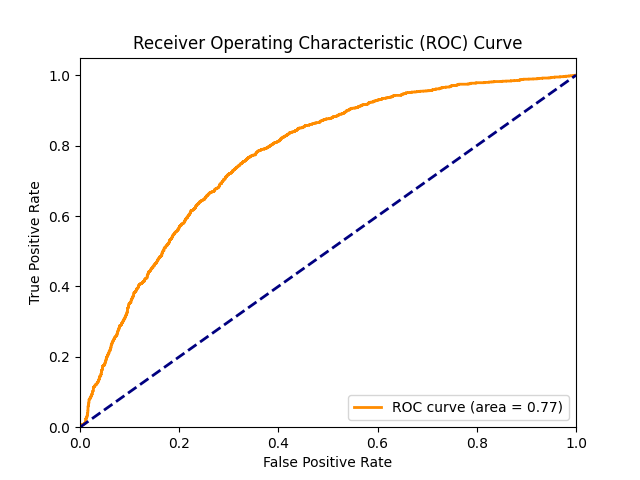}
        \caption{ROC AUC GNN}
    \end{subfigure}
    \hfill 
    \begin{subfigure}[b]{0.45\textwidth}
        \includegraphics[width=\textwidth]{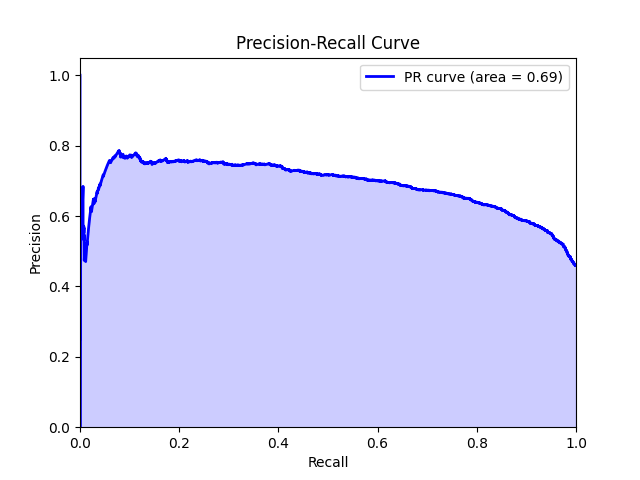}
        \caption{PR AUC GNN}
    \end{subfigure}
    
    \caption{ROC and PR AUC for the Italian model - Testing set}
    \label{fig:rocprit}
\end{figure}

\begin{figure}[htp]
    \centering
    
    \begin{subfigure}[b]{0.45\textwidth}
        \includegraphics[width=\textwidth]{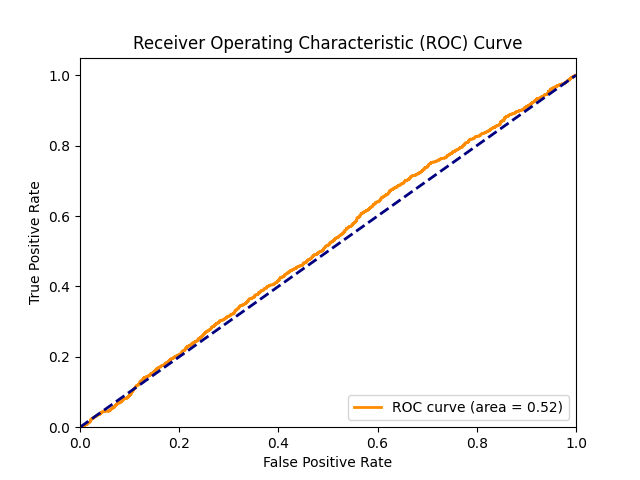}
        \caption{ROC AUC NN}
    \end{subfigure}
    \hfill 
    \begin{subfigure}[b]{0.45\textwidth}
        \includegraphics[width=\textwidth]{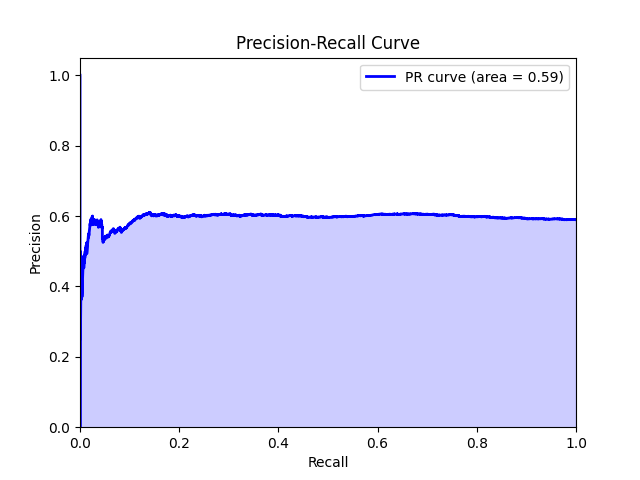 }
        \caption{PR AUC NN}
    \end{subfigure}
    
    \begin{subfigure}[b]{0.45\textwidth}
        \includegraphics[width=\textwidth]{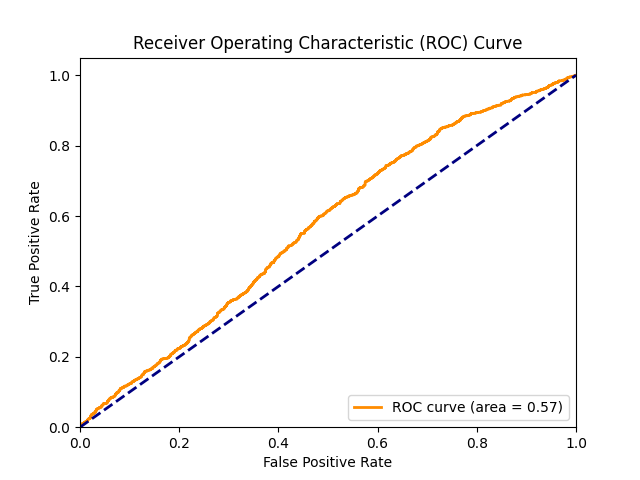}
        \caption{ROC AUC GNN}
    \end{subfigure}
    \hfill 
    \begin{subfigure}[b]{0.45\textwidth}
        \includegraphics[width=\textwidth]{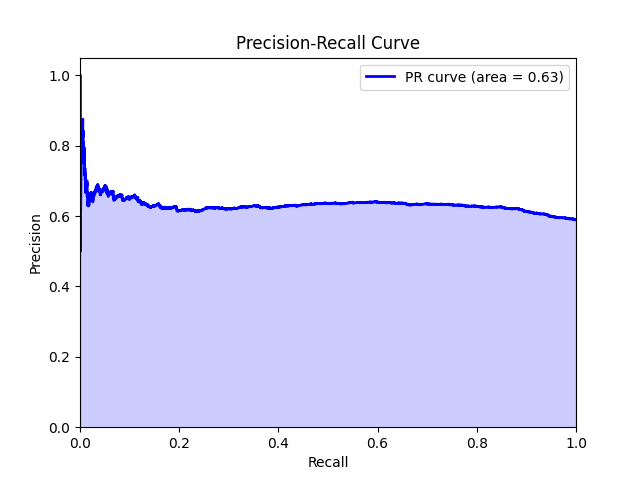}
        \caption{PR AUC GNN}
    \end{subfigure}
    
    \caption{ROC and PR AUC for the St. Gallen and Graubünden model - Testing set}
    \label{fig:rocprsw}
\end{figure}

\begin{figure}[htp]
    \centering
    
    \begin{subfigure}[b]{0.45\textwidth}
        \includegraphics[width=\textwidth]{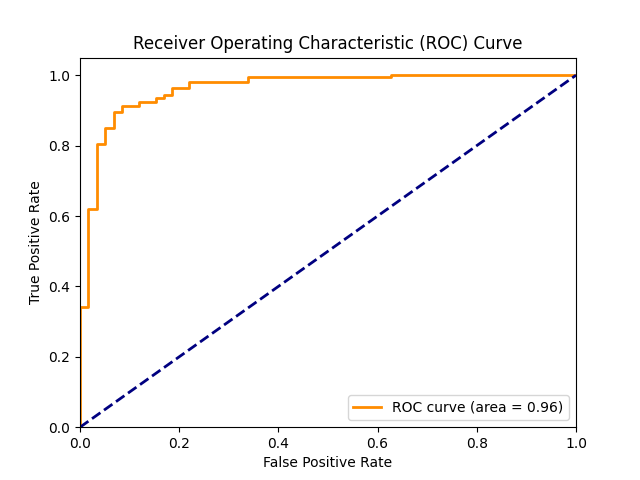}
        \caption{ROC AUC NN}
    \end{subfigure}
    \hfill 
    \begin{subfigure}[b]{0.45\textwidth}
        \includegraphics[width=\textwidth]{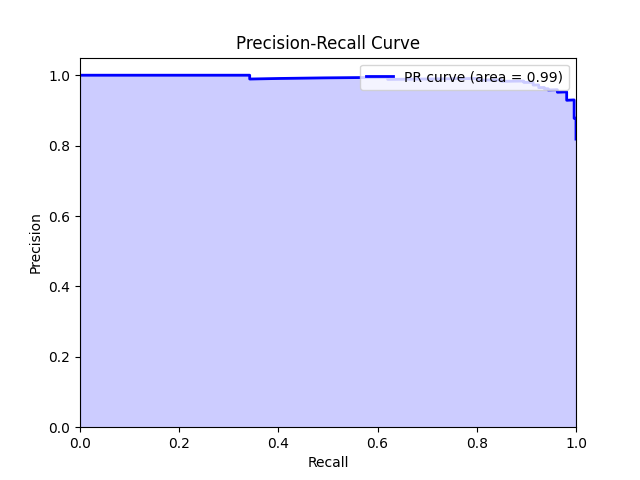 }
        \caption{PR AUC NN}
    \end{subfigure}
    
    \begin{subfigure}[b]{0.45\textwidth}
        \includegraphics[width=\textwidth]{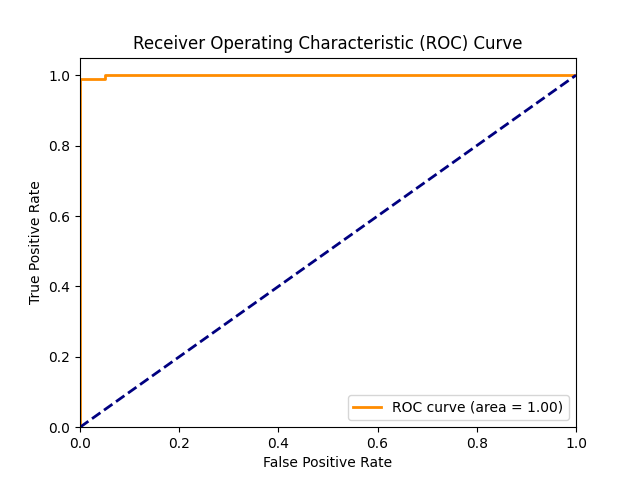}
        \caption{ROC AUC GNN}
    \end{subfigure}
    \hfill 
    \begin{subfigure}[b]{0.45\textwidth}
        \includegraphics[width=\textwidth]{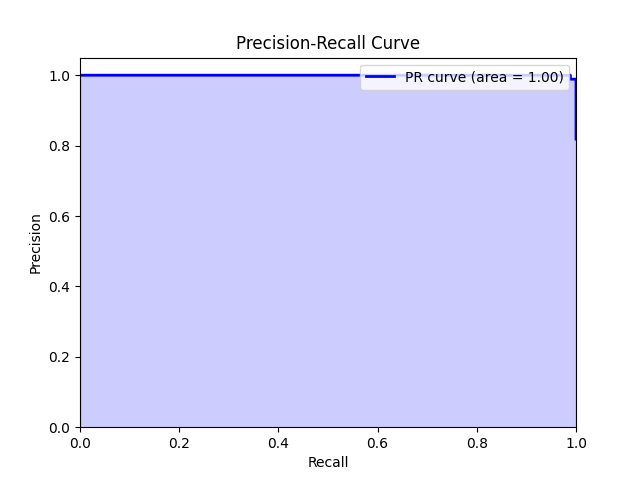}
        \caption{PR AUC GNN}
    \end{subfigure}
    
    \caption{ROC and PR AUC for the Ticino model - Testing set}
    \label{fig:rocprti}
\end{figure}

\begin{figure}[htp]
    \centering
    
    \begin{subfigure}[b]{0.45\textwidth}
        \includegraphics[width=\textwidth]{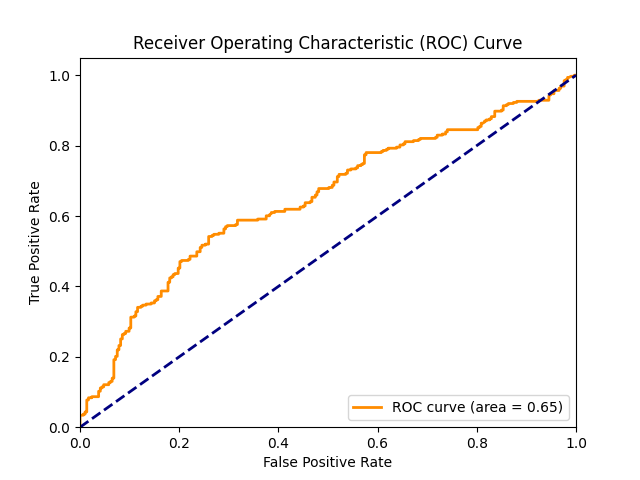}
        \caption{ROC AUC NN}
    \end{subfigure}
    \hfill 
    \begin{subfigure}[b]{0.45\textwidth}
        \includegraphics[width=\textwidth]{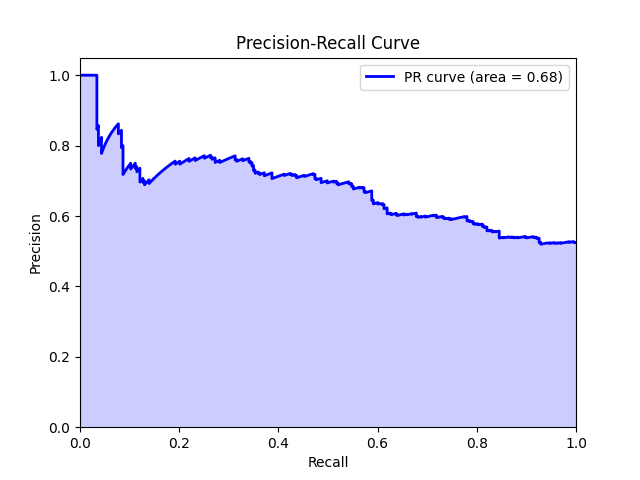 }
        \caption{PR AUC NN}
    \end{subfigure}
    
    \begin{subfigure}[b]{0.45\textwidth}
        \includegraphics[width=\textwidth]{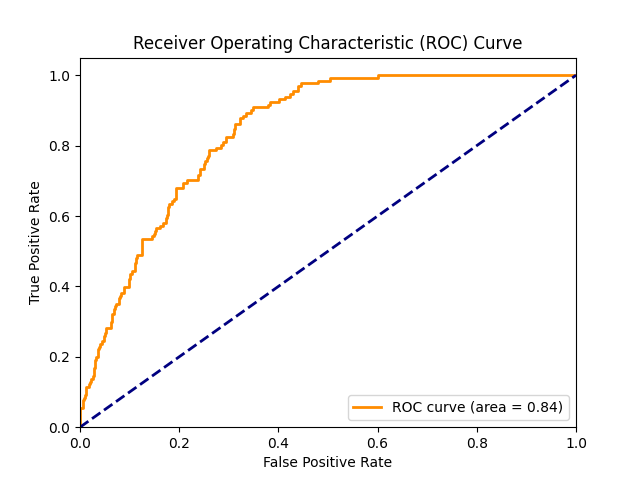}
        \caption{ROC AUC GNN}
    \end{subfigure}
    \hfill 
    \begin{subfigure}[b]{0.45\textwidth}
        \includegraphics[width=\textwidth]{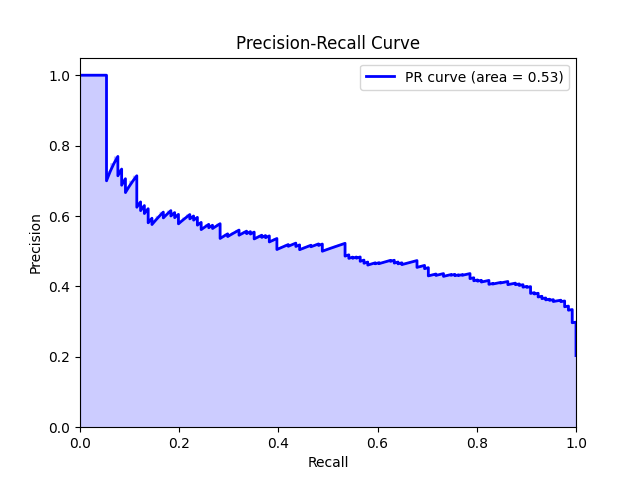}
        \caption{PR AUC GNN}
    \end{subfigure}
    
    \caption{ROC and PR AUC for the US model - Testing set}
    \label{fig:rocprus}
\end{figure}

\end{document}